\documentclass[12pt]{spieman}
\usepackage[utf8]{inputenc}
\usepackage{authblk}
\usepackage[square,sort,comma,numbers]{natbib}
\usepackage{graphicx}
\usepackage{float}
\usepackage{gensymb}
\usepackage{url}
\usepackage{caption}
\usepackage{subcaption}
\usepackage{amssymb}
\usepackage{amsmath}
\usepackage{siunitx}
\usepackage[normalem]{ulem}
\usepackage{wrapfig}

\usepackage{pbox}
\usepackage{lineno}
\title{Fiber positioning in microlens-fiber coupled integral field unit}

\author[a,b,*]{Sabyasachi Chattopadhyay}
\author[a,b,c]{Matthew A. Bershady}
\author[a]{Marsha J. Wolf}
\author[a]{Michael P. Smith}
\author[a]{Andrew Hauser}

\affil[a]{University of Wisconsin, Department of Astronomy, 475 North Charter Street, Madison, WI 53706, USA}

\affil[b]{South African Astronomical Observatory, 1 Observatory Rd, Observatory, Cape Town, 7925, South Africa}

\affil[c]{Department of Astronomy, University of Cape Town, Private Bag X3, Rondebosch 7701, South Africa}

\begin{document}

\maketitle

\begin{abstract}

A generic fiber positioning strategy and a fabrication path are presented for microlens-fiber-coupled integral field units. It is assumed that microlens-produced micro-images are carried to the spectrograph input through step-index, multi-mode fiber, but our results apply to micro-pupil reimaging applications as well. Considered are the performance trades between the filling percentage of the fiber core with the micro-image versus throughput and observing efficiency. A merit function is defined as the product of the transmission efficiency and the \'etendue loss. For a hexagonal packing of spatial elements, the merit function has been found to be maximized to 94\% of an ideal fiber IFU merit value (which has zero transmission loss and does not increase the \'etendue) with a microlens-fiber alignment (centering) tolerance of 1 $\mu$m RMS. The maximum acceptable relative tilt between the fiber and the microlens face has been analyzed through optical modeling and found to be 0.3$\degree$ RMS for input f-ratio slower than f/3.5 but it is much more relaxed for faster beams. From the acceptable tilt, we have deduced a minimum thickness of the fiber holder to be 3 mm for 5$\mu$m clearance in hole diameter relative to the fiber outer diameter.  Several options of fabricating fiber holders have been compared to identify cost-effective solutions that deliver the desired fiber positioning accuracy. 
% Sun-Light Tech
Femto-second laser-drilling methods from commercial vendors deliver holes arrayed on plates with a relative position accuracy of $\pm$1.5 $\mu$m RMS, similar diameter accuracy, and with an aspect ratio of 1:10 (diameter:thickness). 
% Femtoprint
One commercial vendor combines femtosecond laser-drilling with photo-lithographic etching to produce plates with thickness of 5 mm, but with similar ($\pm$1 $\mu$m RMS) positioning accuracy and conical entry ports. Both of these techniques are found to be moderately expensive. A purely photo-lithographic technique performed at WCAM (a facility at the University of Wisconsin, Madison), in tandem with deep reactive ion etching, has been used to produce a repeatable recipe with 100\% yield. Photo-lithography is more precise (0.5 $\mu$m RMS) in terms of hole positioning and similar diameter accuracy (1 $\mu$m RMS) but the plate can only have a thickness of 250 $\mu$m.
 
\end{abstract}


% Include email contact information for corresponding author
{\noindent \footnotesize\textbf{*} \linkable{sabyasachi@saao.ac.za} }

\section{Introduction}

The use of two-dimensional arrays of optical fiber as an integral field unit (IFU) in astronomy has dramatically increased the efficiency of observations for extended sources such as galaxies and Galactic nebulae. Optical fibers provide unprecedented flexibility in terms of instrument development and simplified observation. Early instruments, such as DensePak (\citenum{barden}), used clusters of bare fibers to spatially format the sky and rearrange them into a one-dimensional slit at the spectrograph input. However, even the densest packing of fibers (\citenum{kelz,drory}) 
% (Croom et al. \citenum{SAMI}, Sanchez et al. \citenum{Sanchez}, Bundy et al. \citenum{Bundy}) 
leaves interstitial regions between the fiber cores unsampled, due to the fiber shape, clad and buffer. These gaps lead to fiber arrays with only $\sim$60\% on-sky fill factor. This underfill leads to an `incompleteness' in sampling extended sources (and often under-sampling the point spread function (PSF)), and often requires some sort of mitigation. Non-circular fibers do not significantly alter the situation because the non-circular cores are embedded in circular clad and buffer. The application of buffer stripping and light fusing (\citenum{bryant}) can somewhat improve the on-sky fill factor but at the cost of increased focal-ratio degradation (FRD; see below), which introduces efficiency losses. Successful mitigation strategies have often followed the following two paths.

Historically, missing coverage due to fiber gaps has been mitigated through the dithering of bare-fiber arrays on the sky. Dithering is a process through which the IFU is moved with a slight offset between exposures to cover the interstitial regions. It has long been used in several instruments, such as SparsePak (\citenum{bershady}), PPak (\citenum{kelz}), VIRUS-P and VIRUS-W (\citenum{virusp,virusw}), and MaNGA (\citenum{drory}). Hexagonal close-packing of fibers naturally leads to a 3-point dither pattern which helps obtain integral spatial coverage and resolution on par with the fiber core dimension. Despite these positives (\citenum{law}), analysis of dithered exposure can be challenging when observing conditions change significantly from exposure to exposure. Dithering also demands high telescope offset and guiding precision (better than \SI{0.1}{\arcsecond}), without which the photometric and astrometric calibration can be problematic. Additionally, atmospheric differential refraction (both chromatic and field) can make dithering an inaccurate procedure to implement for wide-field applications over a large range of hour angles (\citenum{law}).

An alternative and elegant solution to filling the fiber gaps is to use a micro-lens array (MLA) or multiple MLAs in front of a fiber array. A well-positioned microlens system will collect and feed the light into the fiber.  Such an optical system produces micro-images or micro-pupils (depending on the optical design) of a continuous focal plane and positions them in a sparse 2D array mimicking the position of the fiber cores in an array. This enables contiguous sampling of the focal plane and PSF, eliminating the need or at least reducing the importance of dithering the IFU and it's associated challenges while retaining the flexibility of fibers to feed spectrographs at locations remote from the telescope focal surface. This approach has been taken, e.g., for the GMOS IFU on the Gemini telescopes (\citenum{gmos}), FLAMES \& VIMOS on VLT (\citenum{flames, vimos}) and more recently for MEGARA on the GTC (\citenum{megara}). In principle, MLAs can provide nearly complete coverage of a target ($\>$99\% with hexagonal MLA and uncovered corners). In practice, the corners often have inferior optical properties and are best masked, but still delivering $\sim$ 90\% integral coverage (\citenum{Yan2013},\citenum{koala}). 

Other advantageous features of MLA coupling concern swapping telescope near- and far-field illumination for scrambling. An optical fiber scrambles the spatial information radially and azimuthally and modulates both near-field and far-field illumination. For a typical astronomical spectrograph, both near-field and far-field illumination are of astronomical importance: The near-field is imaged onto the detector, while the far-field determines the illumination of the optics and hence the aberrations that also contribute to the near-field image pattern (most astronomical spectrographs utilize optical designs that are far from diffraction-limited). Various studies (\citenum{avila2012}, \citenum{spronck}, \citenum{Halverson}), \citenum{smith}) have shown that circular fibers are better at scrambling the far-field than rectangular or octagonal fibers, while non-circular fiber is superior for scrambling the near-field.  The transposition of near- to far-field using lenslets and fiber in series can be used to optimize scrambling both for high-stability systems as well as moving-pupil telescopes such as HET and SALT.

However, MLA coupling to fibers must be implemented with care to avoid significant degradation of instrument performance. \'Etendue -- defined as the product of an optical instrument's collecting area ($A$) and acceptance solid angle ($\Omega$), or grasp ($A\times\Omega$), with the total system throughput ($\eta$) -- is typically considered a measure of an optical system's figure of merit (\citenum{renasmith2002}). In the context of fiber optics, which serve to couple two imaging systems that conserve the area--solid-angle product (e.g., a telescope and a spectrograph), increases in grasp introduced by the fiber at constant throughput either (a) diminish the \'etendue of the overall system by overfilling down-stream optics, making for a lossier system; or (b) reduce the spectral resolution by increasing the entrance aperture. Throughput losses introduced by the fiber are additional losses to the system \'etendue. All of these cases can be viewed as a loss of information, and hence an increase in optical entropy.  A fiber optic's grasp is defined as the product of the illuminated cross-sectional fiber-core area and the related solid angle of the illuminating beam, while the total throughput is the product of surface losses and bulk transmission. In the context of MLA coupling to fibers, it is critical to consider how this coupling may lead to increasing the grasp or reducing the throughput delivered by the fiber. 

For example, MLAs offer an advantage of allowing for the fiber input f-ratio to be modulated in order to minimize FRD but have the potential disadvantage that misalignment between MLAs and fiber cores can lead to light loss and geometric FRD (gFRD). FRD is a phenomenon related to fibers in which the input beam f-ratio becomes faster when exiting the fiber. There are several explanations of FRD that include, but are not limited to, fiber-polishing imperfections (\citenum{arthur}), end stress (\citenum{alsmith}), and micro bends (\citenum{carasco}). It is well understood that FRD cannot be entirely eliminated but can only be minimized through certain procedures. FRD may also introduce lossy modes leading to lower transmission. Geometric FRD is introduced due to the misalignment of the input beam with the fiber optical axis and thus geometric in nature (\citenum{wynne}); gFRD can be eliminated with proper optical design and alignment. Light-loss due to misalignment can be mitigated by underfilling the fiber cores with the MLA micro-images, but this comes at a cost in entropy: Fibers produce radial scrambling in the near-field, which, in the case of an under-filled fiber entrance aperture, tends to lead to a more filled fiber exit aperture (this radial scrambling is a function of fiber properties, including length, e.g. Figure 8 in \citenum{yan2016}). This last point is the crux of the challenge with MLA-fiber coupling. 

Ren \& Allington-Smith (\citenum{renasmith2002}) have explored in detail several of these key issues of fiber-microlens coupling. For example, they consider the important performance implications of under or overfilling the fiber input surface, including the effects of spherical aberration and diffraction from the microlens. In general, they conclude the use of microlenses requires an oversize factor between the fiber core and microlens clear aperture, an equivalent of the mechanical aperture. This oversize factor impacts the effective slit width seen by the spectrograph (pseudo-slit), and hence lowers spectral resolution at constant collimator focal-length and grating dispersion. However, their analysis is based on the use of plano-convex micro-lenses at the input and output of the fiber. This optical configuration is mechanically simple to implement and serves to transfer a micro-pupil image to and from the input and output fiber faces, respectively. As they note, this transfer implementation introduces non-telecentricity to the micro-pupil image, and hence introduces gFRD due to the azimuthal scrambling properties of the fibers even in the case of perfect fiber-microlens alignment. Their solution to this form of gFRD places limits on the input beam speed to the microlens to low values that are not optimal for other design considerations. They also identify an optimum input f-ratio and fiber core size based on the micro-bend theory for FRD that has not been confirmed in practice. In our study, we extend their work to provide (i) a more generic treatment of the impact of fiber alignment and fill factor with respect to an optically defined chief ray for both micro-pupil and micro-image transfer system; (ii) an explicit expectation that if the fiber input near-field is under-filled, the output end will be filled with near uniformity (while in detail the output near-field illumination will depend on fiber length, input fill factor and centricity of the illumination [\citenum{Yan_2015}]); and (iii) practical limits on lenslet radius of curvature (RoC). The latter is critical since it places constraints on the input beam speed to the lenslets, which connects back to the issue of geometric FRD in the context of lenslets used as pupil reimagers.

\begin{figure}[H]
\centering
\includegraphics[width = 0.7\linewidth]{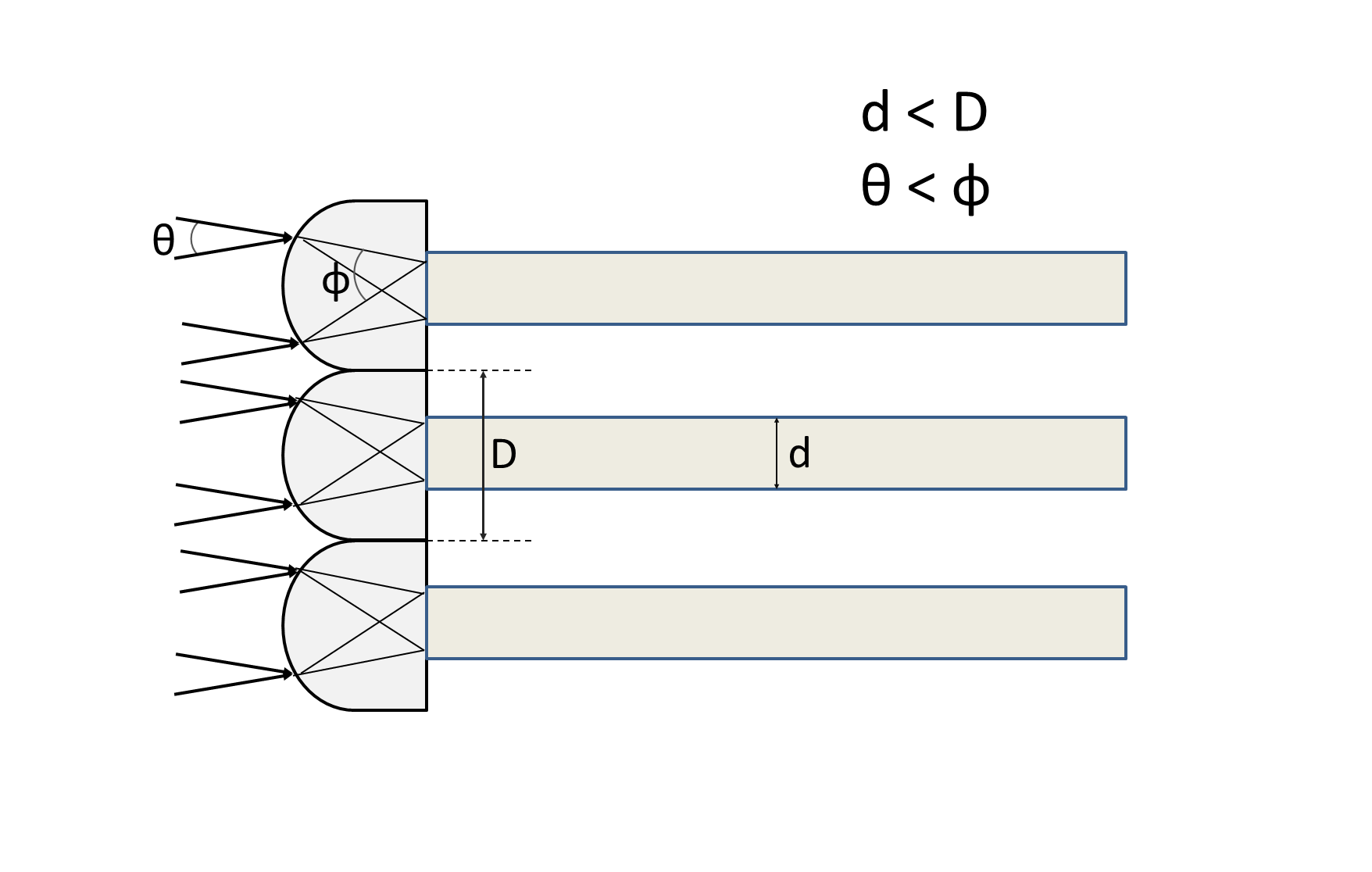}
\caption{Figure depicting the requirement of a sparsed fiber array for a microlens-fiber coupled integral field unit (IFU). The fiber diameter (d) is smaller than microlens (D) as the microlens requires a slower beam ($\theta$) to be fed to be manufacturable while the fiber needs a faster beam ($\phi$) to minimize the focal ratio degradation (FRD). Since the figure is presented to demonstrate the requirement of sparse distribution of fiber for a microlens feed IFU, please note that the non telecentric beam entering the fiber as shown in the figure would increase effective $\phi$ and in turn will increase FRD from its nominal value.}
\label{fig:sparse}
\end{figure}

In our framework we assume MLAs must feed telecentric images (pupil or near-field) to fibers at fast f-ratios ($\sim 3$) to minimize the effect of FRD. On the other hand, the microlenses need to be fed at a slow f-ratios to ensure that their radius of curvature is manufacturable. This difference in beam speeds, and in turn angular area ($\Omega$), leads to a difference in surface area between the MLA and fibers, with fibers requiring lesser area than MLA. Thus the fiber array must be sparsely populated yet precisely matching the position of the micro-pupil/image array produced by the MLA. Positioning  fibers as shown in Figure \ref{fig:sparse} leads to positioning tolerance generated from position inaccuracy as well as diameter inaccuracy. This tolerance will affect the overall throughput of the system.

Consequently, the success of using MLA's in front of a fiber array in an IFU depends not only on the relative positioning of fibers and microlenses but also on the amount of the fiber core that is filled as well as minimization of tip/tilt error between the fiber and the microlens surface. In this paper, we will define merit functions that quantify this `success,' and discuss the strategy to optimize the IFU design in section \ref{sec:strategy}. Section \ref{sec:lithography} describes one of the methods to implement the required fiber positioning. In section \ref{sec:results} we summarize our findings.

\section{Micro-lens coupled fiber IFU design strategy}
\label{sec:strategy}

The key to developing a fiber-MLA coupled IFU is to optimize the trade-offs between throughput loss and increase in grasp such that optical entropy increase is minimized. For the purposes of this development, we will consider a system in which the FRD and transmission properties due to the fiber alone to be fixed. We first isolate the trade-offs between the fraction of the fiber optical core that is filled by the MLA micro-image and MLA--fiber centering precision. In Section~\ref{subsec:thickness} we consider the impact of gFRD on the requirements for fiber angular alignment.  

Ideally, to achieve minimum entropy gain on the one hand it is desirable to fill the entire fiber entrance core since radial scrambling will tend to fill the exit core regardless of the entrance fill factor. However, in this fully-filled input core scenario, the positioning accuracy of the micro-image onto the fibers (or, alternatively, the fiber location) must be extremely high to minimize throughput loss at the fiber-microlens junction. Such high precision position requirements can dominate the cost of fiber-based IFU development.

The minimum entropy gain also depends on the encircled energy within the fiber core for a given micro-image diameter. If we define EE99 as the radius of 99\% encircled energy, clearly the quality of the microlens design should deliver a micro-image that is at least as small as the fiber core radius.

Given these two considerations, it is important to determine the fraction of the fiber core to be filled by the micro-image given the achievable alignment precision between fiber and MLA. Hence a merit function is defined to find the exact fiber core fill fraction that includes a decenter model of the fiber array holder and encircled energy distribution within the micro-image. We define the fractional change in grasp due to the fiber coupling as:
\begin{equation}
\label{eq:etendue}
E(A,\Omega) = \frac{A_o\Omega_o}{A_i\Omega_i},
\end{equation} 
where $A_i$ and  $A_o$ are input and output micro-image areas, respectively, at the fiber face while $\Omega_i$ and $\Omega_o$ are the respective solid angles of the beams. For a microlens-fiber throughput of $\eta$ the merit function of a single spaxel design is then defined as:
\begin{equation}
\label{eq:singlemf}
M_s = \frac{\eta}{E(A,\Omega)} = 
% \frac{\eta}{\frac{A_o\Omega_o}{A_i\Omega_i}} = 
\frac{\eta{A_i\Omega_i}}{A_o\Omega_o}.
\end{equation} 

For a typical multi-mode fiber used in astronomy, $A_o$ remains fixed and equal to the fiber core size as the beam completely covers the fiber output face. Consequently, $M_s$ increases for an increase in input micro-image size. $M_s$ also increases with increasing $\Omega_i$ since FRD modulates slower input beams more compared to the faster beams; while the ratio of $\Omega_i/\Omega_o$ remains less than unity due to FRD, it approaches unity as the input beam approaches the fiber numerical aperture. (It is assumed in all of this discussion that the fiber input beam does not exceed the numerical aperture since this would be a lossy application.) Putting these factors together, we expect $M_s$ to be less than unity in practice. Our objective is to maximize $M_s$ by careful design of a microlens system that maximizes $\Omega_i$ for a suitable choice of $A_i$ within the achievable fiber positioning accuracy. 

As long as the fiber positioning accuracy does not depend on the fiber spacing, $\Omega_i$ and $A_i$ can be decoupled so that we may consider the optimum $A_i/A_o$ ratio in the context of the positioning accuracy alone. For an appropriate statistical treatment let $F(p,\mu,\sigma)$ be a gaussian decenter distribution of fiber position $p$ with mean $\mu$ and standard deviation $\sigma$. The average of single spaxel merit function convolved with the decenter distribution is defined as the IFU merit function $M_i$ described as,
\begin{equation}
\label{eq:finalmf}
M_i = <M_s\circledast{F(p,\mu,\sigma)}> = <{\frac{\eta{A_i\Omega_i}}{A_o\Omega_o}}\circledast{F(p,\mu,\sigma)}>.
\end{equation} 
Maximization of $M_i$ may be achieved through microlens-fiber optical design and ensuring relative positioning accuracy as described by Perez-Calpena et al. (\citenum{perez}) for MEGARA in GTC. The study treated the microlens positioning error within lenslet array and microlens to fiber positioning error as separate entity. Here $F(p,\mu,\sigma)$ is a combination of both of these errors. The metric has been used in Section \ref{subsec:corefill} in finding the best fiber holder technique among the technologies available while defining the requirements for the optical design at the same time. 

\subsection{Limit of input solid angle}
\begin{figure}[H]
\centering
\includegraphics[width = 0.8\linewidth]{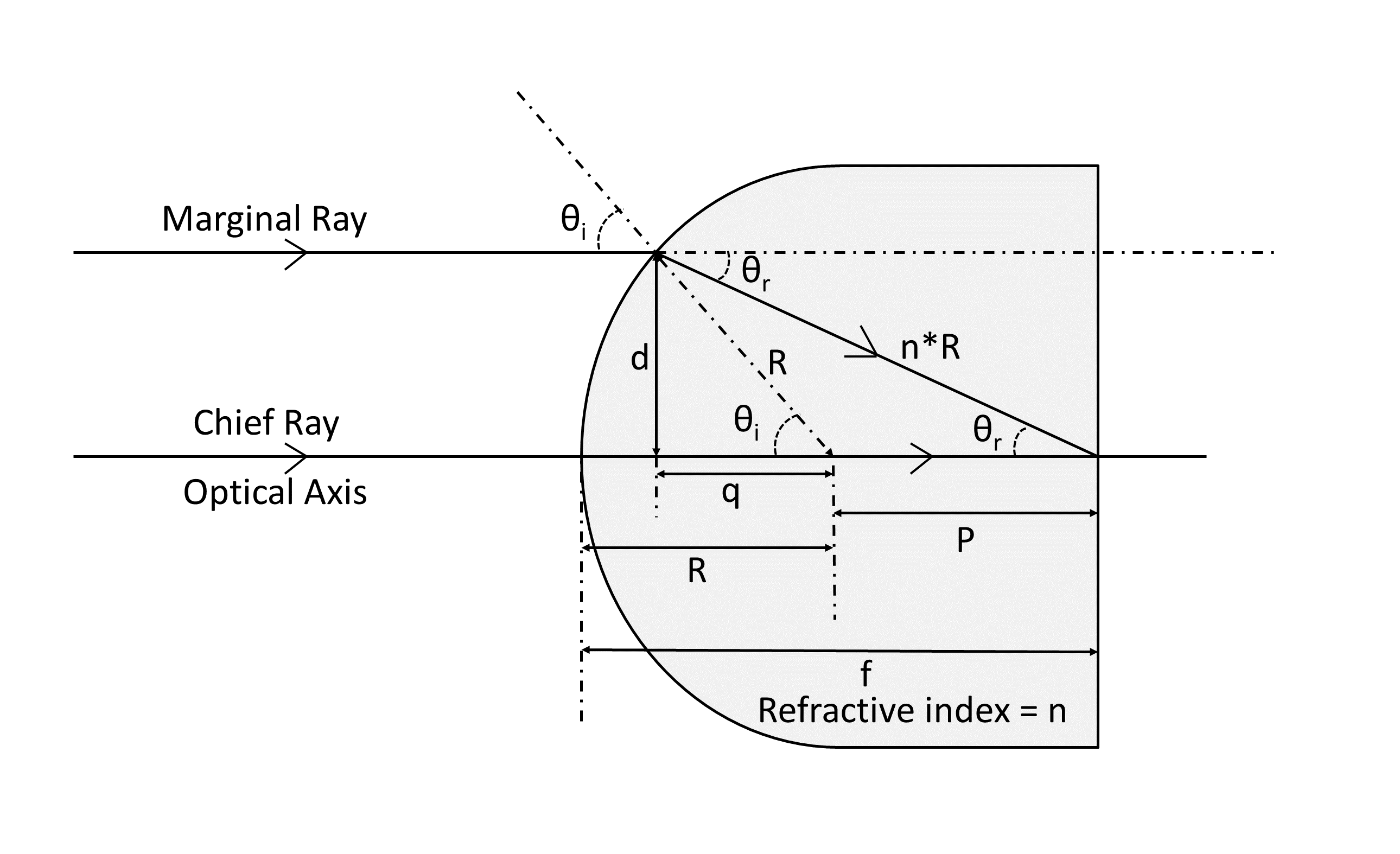}
\caption{Effect of aperture on the upper limit of the input solid angle.}
\label{fig:RoC}
\end{figure}

Equation \ref{eq:finalmf} suggests that to maximize the merit function, we must maximize the input solid angle $\Omega_i$ which would minimize the effects of FRD. Ideally, the numerical aperture (NA) of the fiber (typically 0.22) would put the upper limit on the $\Omega_i$ for bare fiber IFU. However, with microlens, the limit on $\Omega_i$ is defined by the requirement of lenslet clear aperture. We describe this through the figure shown in Figure \ref{fig:RoC}. The marginal ray of a collimated beam is going to pass through a plano-convex microlens and produce a micro-pupil/image at the back of the lenslet. We have considered a plano-convex lens just to visualize the scenario. Here, R, d, $\theta_i$, $\theta_r$, n, and f are the radius of curvature, semi-diameter of the clear aperture, input angle, refracted angle, refractive index, and lenslet focal length respectively.

Since
\begin{equation}
\sin\theta_r = \frac{d}{nR},
\end{equation}
we have
\begin{equation}
\tan\theta_r = \frac{\sin\theta_r}{\cos\theta_r} = \frac{\frac{d}{nR}}{\sqrt{1-({\frac{d}{nR}})^2}}.
\end{equation}
Also,
\begin{equation}
\tan\theta_r = \frac{d}{P+q}.
\end{equation}
Combining equation 5 and 6 we find:
\begin{equation}
P = \sqrt{({nR})^2 - d^2} - \sqrt{R^2 - d^2}.
\end{equation}

The criteria for an acceptable clear aperture is: Within the clear aperture, all the rays of a collimated beam parallel to the optical axis of the lenslet should pass through the focus. To ensure the refracted light must pass through the focus at the focal plane which is the backplane of the lenslet, this condition must be met:
\begin{equation}
P \leq f-R.
\end{equation}
For a plano convex lens using the lens-makers formula we have
\begin{equation}
R = (n-1)f.
\end{equation}
Combining 8,9:
\begin{equation}
P \leq (\frac{2-n}{n-1})R.
\end{equation}
Finally, from 7 and 10:
\begin{equation}
\frac{2-n}{n-1} \geq \sqrt{n^2 - {(\frac{d}{R}})^2} - \sqrt{1 - ({\frac{d}{R}})^2}.
\end{equation}

Equation 11 shows that, for the refractive index of fused silica, the value of d can be as high as 0.65$\times$R. From equation 9, a limit on R would put a limit on the focal length of the microlens. Again, for fused silica, d is can only be as high as 29\% of the focal length. This limit falls nicely along with the circular microlens profile which is within the regime of a parabolic surface at d$=$0.62$\times$R and hence the spherical aberration is minimal. Even the remaining aberration is taken out by now-standard manufacturing processes that generate a refractive index profile decreasing from the lenslet center to the edge (Section 7.7 of Chapter 7 from \citenum{handbook}). The manufacturing limit of the clear aperture is defined by the thickness of the substrate which primarily defines the shortest possible focal length by limiting the radius of curvature. At the limit of d $\sim 0.65 \times$R, the numerical aperture (shag over semi-diameter of clear aperture) gets limited to 0.22 which also matches nicely with the NA of a typical multi-mode optical fiber. However, as described, the acceptable NA is dependent on the focal length and thus substrate thickness. Small substrate thickness ($\sim 100 \mu$m) would demand a limiting NA of 0.14 only. For a given fiber core size and telescope f-ratio, theoretically, the microlens focal length and practically the substrate thickness acts as a link between the clear aperture and the maximum attainable $\Omega_i$ (Chattopadhyay et al. in preparation). 

\subsection{Slit-Mask Integral Field Unit for SALT}
\label{subsec:inst_config}

As an example of its generic description, we have implemented, analyzed and discussed our theoretical merit function for the IFU optomechanical design of the Slit-Mask IFU (SMI), an instrument for the South African Large Telescope (SALT). As a pupil scrambling IFU, the current SMI design deploys 270 spaxels (spatially contiguous fiber-microlens elements) at the telescope science focal plane. The spaxels are hexagonal and each covers $\sim$1.33" (corner to corner) on the sky. Physically, each hexagonal microlens is $\sim$370 $\mu$m wide, transmits an f/4.2 telescope beam as a part of the focal plane at the input of the fiber as a micro-pupil. SMI is expected to use 200 $\mu$m core fibers. The fibers must be positioned at the back of microlens in a way that the fiber core engulfs the entire micro-pupil to avoid any throughput loss. However, it is also important to explore the pros and cons of overfilling the fiber core as well. It is possible to prepare the fiber array and reflect their positions on the MLA micro-pupil pattern. The custom positioning of microlenses in an MLA is significantly costlier than maneuvering the fiber positions. To ensure fiber positioning is accurate enough to minimize etendue loss, we break out the requirements into three components:

\begin{enumerate}
    \item fiber core centers must be aligned with micro-image centers (positioning accuracy);
    \item hole diameters should be as close as possible to the fiber outer diameter (including clad and buffer; diameter accuracy);
    \item holes should be deep enough to limit tilt at the fiber face relative to microlens face to within the acceptable tolerance for non-telecentricity (hole depth).
\end{enumerate}

Typically, fibers are plugged into holes in a plate, which we refer to as a fiber holder. The holder hole pattern, therefore, mimics the micro-pupil (pupil scrambling IFU) or micro-image (image scrambling IFU) pattern.

\subsection{Angular (tilt) alignment requirement}
\label{subsec:thickness}

The design and fabrication of a fiber holder depend on the requirements for sky-fiber positioning accuracy as well as spectrograph acceptable fiber output. The positioning accuracy is driven by the micro-images produced by the MLA. However, the tip/tilt in the coupling face between a fiber and a microlens also defines the fiber output beam degradation. This kind of FRD is known as geometric FRD (gFRD) which is introduced due to misalignment of the input beam with the fiber optical axis and thus geometric in nature (\citenum{wynne}). The acceptable limit of gFRD drives the tip/tilt tolerance between microlens and fiber. 

\begin{figure}[H]
\centering
\includegraphics[width =0.8\linewidth]{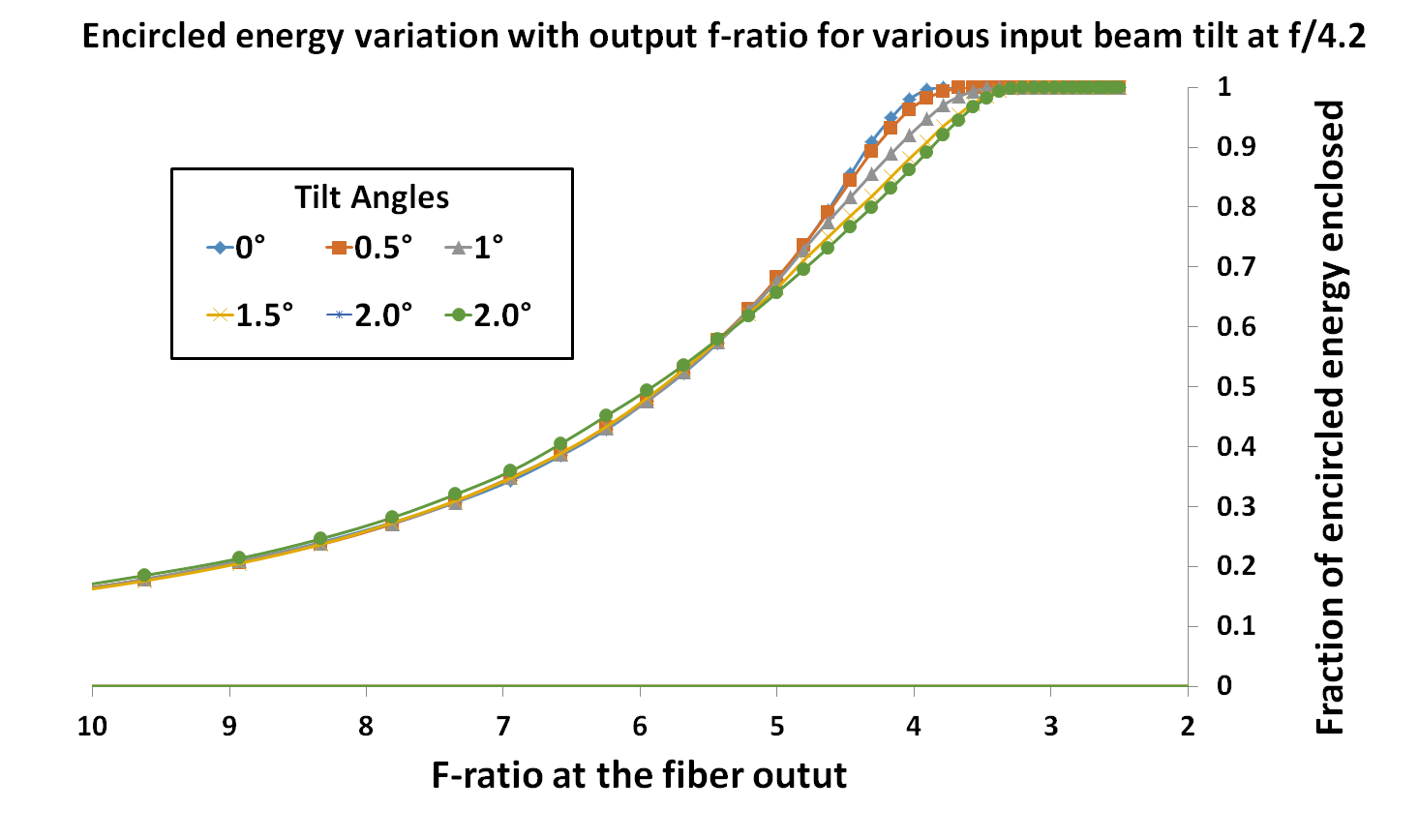}
\caption{Encircled energy variation with output f-ratio for a 1~m fiber fed with a f/4.2 input beam and non-telecentric injection varying from $0^\circ$ to $2^\circ$ at a step of $0.1^\circ$. Color of the curves different tilt angles in degree between fiber and microlens face.}
\label{fig:EEvsFratio}
\end{figure}

\begin{figure}[H]
\centering
\includegraphics[width = 0.8\linewidth]{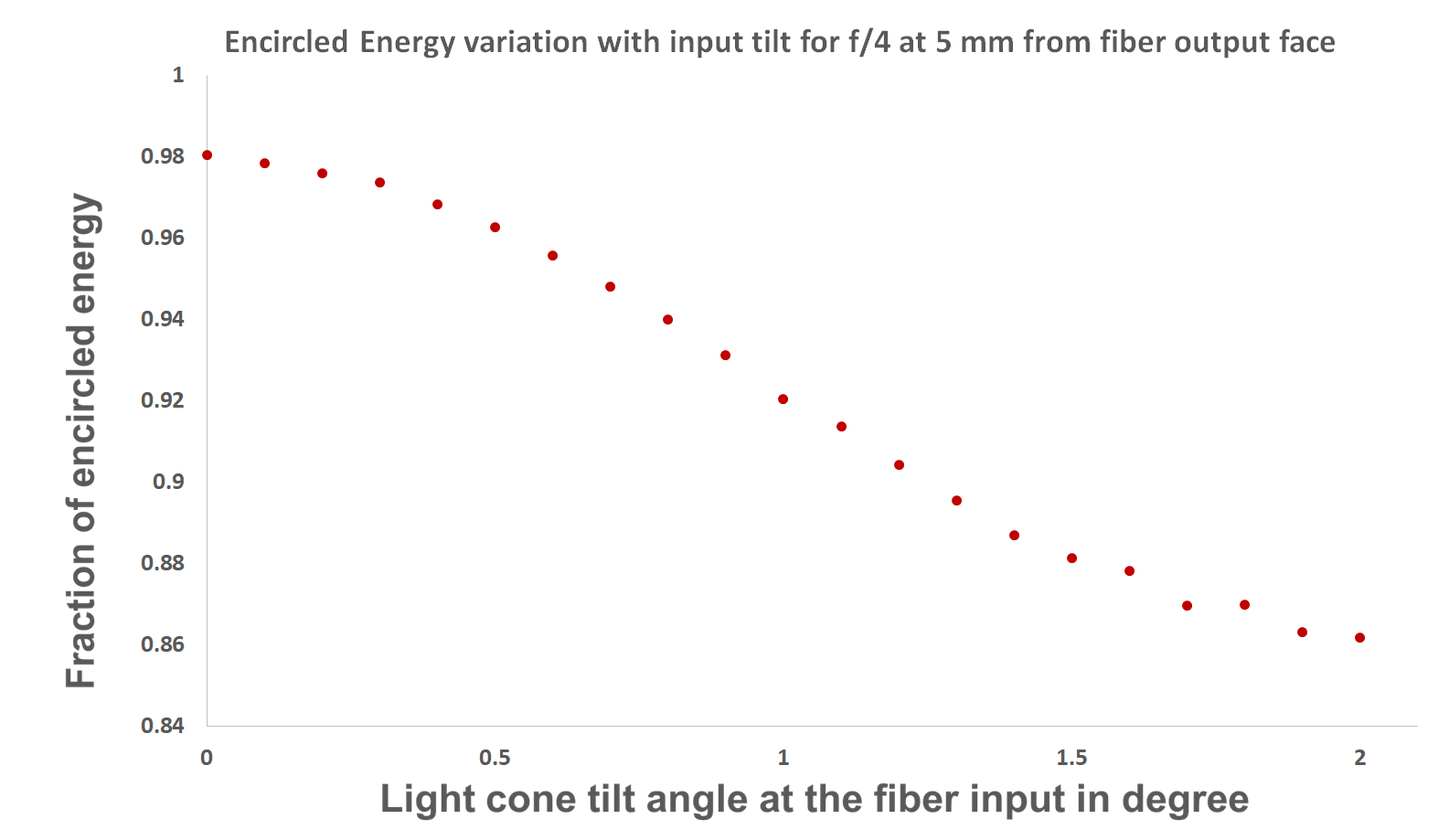}
\caption{Fiber output encircled energy (EE) at f/4 as a function of non-telecentric injection (tilt), as measure at a distance of 5 mm from fiber output face. Near-field effects due to the finite size of the fiber output face are 2\% as reckoned from EE at zero tilt angle. }
\label{fig:EEvsTilt}
\end{figure}

We have performed a Zemax\textcopyright\ simulation to understand the effect of tip and tilt on the energy distribution at the fiber output for our example application of a SALT slit-mask IFU. We have chosen the input beam speed to be f/4.2 which is the telescope beam speed. The practical FRD measurement defines the expected EE98 for the purpose of simulation which is f/4 in our case. In the simulation, a 1~m long fiber of 300 $\mu$m core is fed with an f/4.2 beam and the fiber output is captured on a surface at a distance of 5 mm from the fiber output face. At this surface, the extended source encircled energy distribution is simulated using 5 million rays. The pupil of the optics does not coincide with the plane at 5 mm. While we have not calculated to what degree the finite size of the near field image impacts the encircled energy at 5mm, this is unimportant for our purposes since all we care about is the degradation relative to no tilt, i.e., roughly 98\% encircled energy as seen at the y-intercept. The input beam tilt (equivalent to the tilt between MLA and fiber) is varied from $0^\circ$ to $2^\circ$ at a step of $0.1^\circ$. The EE distribution is plotted against the output f-ratio in Figure \ref{fig:EEvsFratio} for different tilt angles. From this distribution, an EE variation versus tilt is plotted for the output f-ratio of f/4 in Figure \ref{fig:EEvsTilt}. Until $0.3^\circ$ the EE does not vary significantly. However, there could be tilt introduced from other sources (e.g., MLA defects, global MLA misalignment, etc.) so we only allow a fraction (33\% = $0.1^\circ$) of the total available budget to be attributed to the fiber angular misalignment. This acceptable tilt angle can be converted into the fiber-holder thickness depending on the difference between fiber diameter and fiber holder hole diameter. For example, to accept a 5 $\mu$m larger hole diameter than the fiber diameter (and hence 5$\mu$m decenter error), the fiber holder thickness must be $\sim$3 mm. 

\section{Fiber holder development}
\label{sec:lithography}
\subsection{Fiber holder development techniques}
\label{subsec:schemes}

We have explored three technologies to develop fiber holders with the aim of determining the accuracy and precision achievable by these different techniques in relative hole position, diameter, and achievable hole depth. The three techniques are laser drilling, photo-lithography, and a combination of both.

\begin{figure}[H]
\centering
  \includegraphics[width=0.8\linewidth]{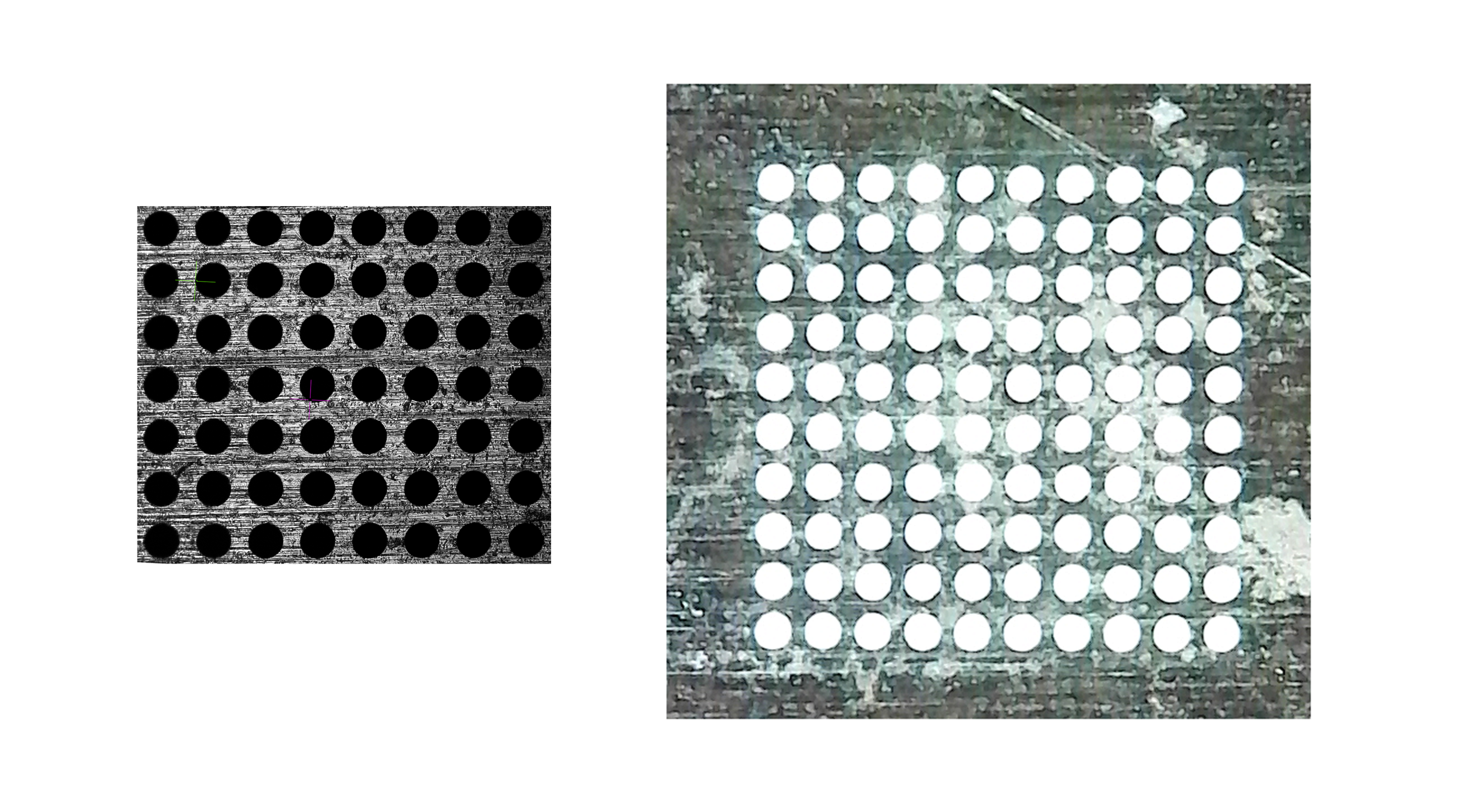}
  \caption{Back and front illuminated image of Sun-Light Tech produced 10$\times$10 hole array of 100 $\mu$m holes with a pitch of 150 $\mu$m. The RMS positioning and diameter error was found to be $\pm$ 1.5 $\mu$m measured by using microscopic calibrated image.}
  \label{fig:slt}
\end{figure}

\begin{figure}[H]
  \centering
  \includegraphics[width=0.6\linewidth]{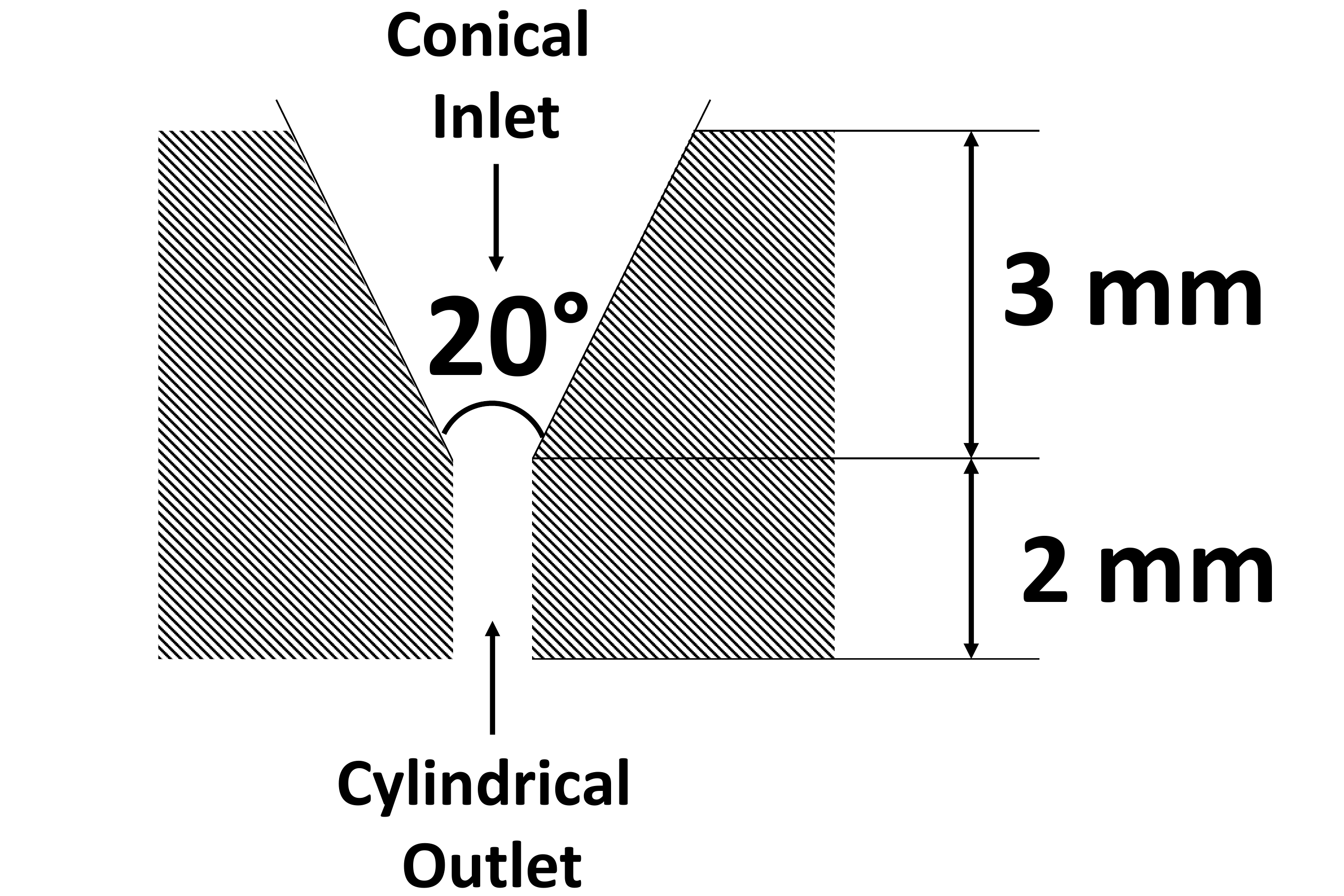}
  \caption{FemtoPrint design of hole as fiber holder that has a conical entry to the desired cylindrical hole which demands sparse fiber positioning.}
  \label{fig:fp}
\end{figure}

\begin{itemize}
    \item \textbf{Sun-Light Tech\textcopyright\ (SLT)} uses a femtosecond-pulsed laser to drill metal or glass plates with an aspect ratio of 10:1 between hole depth and diameter. For a hole diameter of 280 $\mu$m (5 $\mu$m larger than the fiber outer diameter including buffer for SMI), we can achieve a hole depth of $\sim$ 2.8 mm. We have inspected an 1 mm thick 10$\times$10 rectangular array of 100 $\mu$m diameter holes spaced at 150 $\mu$m center to center (refer to Figure \ref{fig:slt}). The microscopic image was used to measure the hole position accuracy and hole diameter accuracy and it is found that both are better than 1.5 $\mu$m RMS. 
    
    \item \textbf{Wisconsin Center for Advanced Microelectronics (WCAM) currently Nanoscale Fabrication Center (NFC)} at the University of Wisconsin, Madison provides an ideal, in-house laboratory facility to perform photolithography. Photolithography is a chemical process that transfers patterns to a metal or glass plate. It has a much higher aspect ratio for hole depth to diameter (150:1) compared to laser-drilled technique, as well as higher accuracy in hole positioning ($\sim$0.5 $\mu$m) and diameter ($\sim$0.5 $\mu$m). We return to this technique in section \ref{sec:lithography}.
    
    \item \textbf{FemtoPrint\textcopyright\ (FP)} is a Swiss manufacturer that uses a combination of the above two techniques to produce a fiber holder. A key feature of FP-produced holders is a conical hole at the entry surface which meets the expected cylindrical hole as shown in Figure \ref{fig:fp}. FP-produced holders are expected to have 2 and 7 $\mu$m RMS hole diameter accuracy for 1 mm and 5 mm thick holders, respectively. In both cases, the hole positioning accuracy for the cylindrical section is 5 $\mu$m RMS. Currently, FemtoPrint is capable of delivering $\pm$ 1 $\mu$m hole position and diameter accuracy over a holder thickness of 5 mm. We term this capability/distribution as modified FP.
\end{itemize}

\subsection{Lithography as a tool for pattern transfer} 
\label{subsec:pattern_transfer}

In this subsection, we describe the fabrication of a fiber holder through photo-lithography. We have used the facilities at WCAM for this purpose. Wet photo-lithography has previously been used (\citenum{sabyasachilitho})  to develop microlens-fiber coupled IFU for the Devasthal Optical Telescope Integral Field Spectrograph (DOTIFS). However, it has been shown that wet photo-lithography can have issues of irregular hole diameter, low etchable thickness (maximum of 100$\mu$m), low throughput ($\sim$ 30\%), etc. In dry photo-lithography, it is possible to use a much thicker substrate, such as a Silicon wafer. We have used 250 and 500 $\mu$m  dual-side polished (DSP) silicon wafers of $<$100$>$ crystal lattice orientation as our substrate. The DSP feature helped us maintain etching uniformity and hole isotropy while the $<$100$>$ orientation is found to be easier to etch with the existing apparatus than other orientations. 

\begin{figure}[h]
\centering
\includegraphics[width = \linewidth]{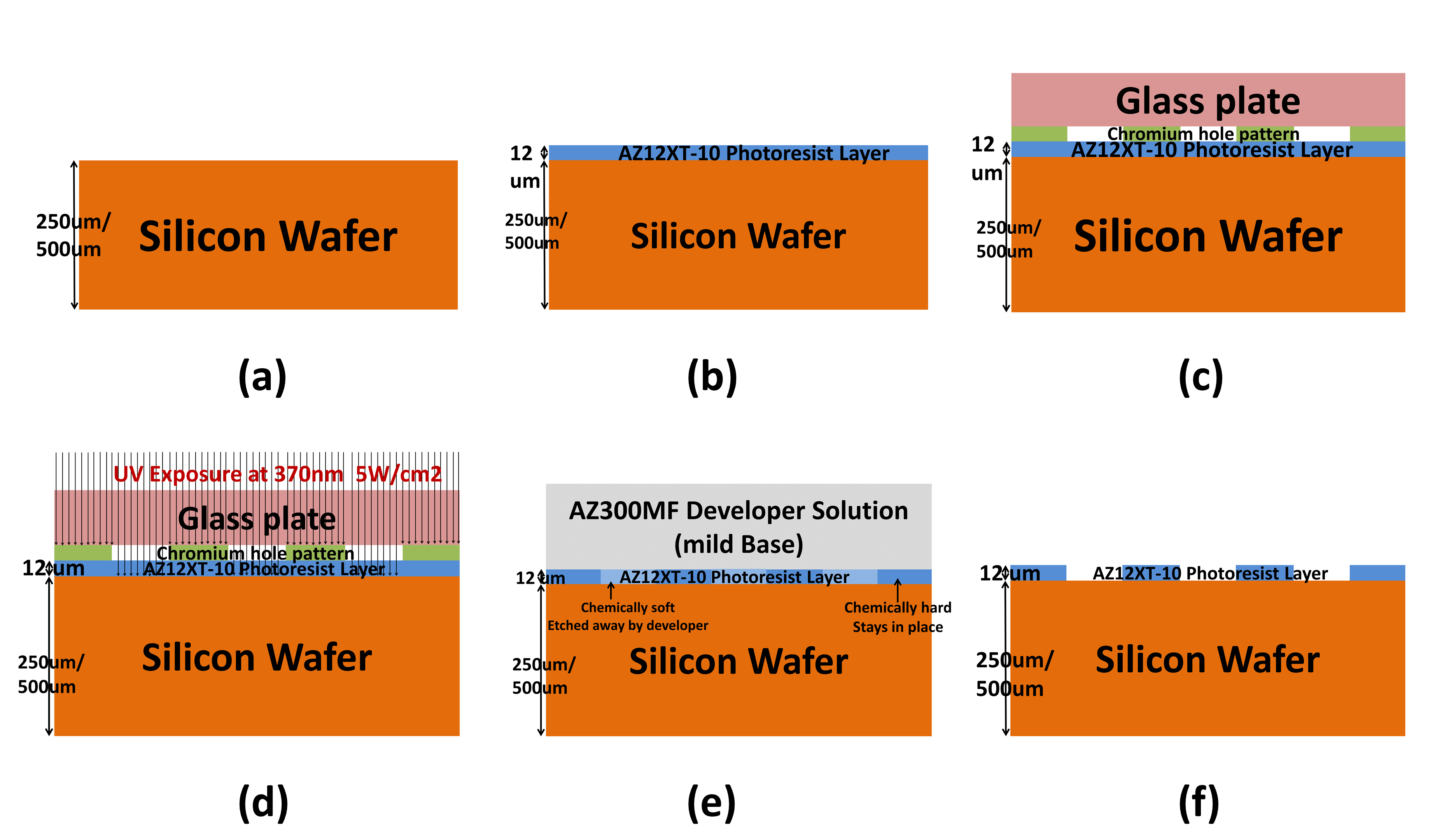}
\caption{Schematic of different steps involved in photo-lithographic procedure used towards developing a fiber holder at WCAM.}
\label{fig:litho}
\end{figure}

Photo-lithography is a chemical technique for precisely etching patterns on a silicon wafer. The positional accuracy of etching patterns can be controlled to a tenth of a micron using ultraviolet (UV) light of 365-370 nm. For prototype development purposes, we have tried to fabricate a fiber holder that can hold one of the smaller size multi-mode fibers ($\sim$100 $\mu$m core, 145 $\mu$m including clad, jacket/buffer) usable for astronomical purpose. We believe if the method can hold position and diameter accuracy for smaller diameter and separation, it would be expected to perform the same for wider fibers with larger separation. Given our requirement for total hole depth and the limitations of silicon wafer thickness for etching, several wafers need to be fabricated and stacked to achieve a thickness of 3 mm. This requirement has an impact on our etching design. Figure \ref{fig:litho} illustrates the photo-lithographic process. We describe the process below with respect to this figure.

\begin{enumerate}

    \item Photo-lithography requires the purest form of substrate. Impurities, if any, can act as a constructive or a destructive catalyst in the etching process leading to a higher or lower etching rate, and hence hole size and regularity within the stipulated etching time. Such impurities thus reduce the yield of acceptable arrays in a wafer. The wafer should also be easily polishable once bonded to the fibers. Since the fiber core and cladding are made of fused silica, silicon is used as the wafer substrate. The chosen silicon wafer has a low percentage ($<$0.001\%) of impurities. The glue, the wafer, and the fiber all have very similar thermal expansion properties and hence will not create any stress-induced internal mechanical issues (insofar as they are properly mounted) in observatory conditions where the temperature can vary from -10\degree to 25\degree.
    
    \item The two factors that determine wafer dimensions are the number of masks that we want to produce at a time and the UV-illumination area. Six masks are fabricated in a single process which is sufficient for achieving a stack thickness of 3 mm. (There can be added thickness of a few 10s of $\mu$m by layers of glue between each pair of wafers.) The corner to corner dimension of a hexagonal IFU holder is 9.4 mm. We have kept an additional 5.6 mm space on all four sides for ease of handling. On the other hand, the ultraviolet (UV) lamp assembly can illuminate an area corresponding to a circle of diameter 75 mm. The wafer dimensions are large enough to create a $2\times3$ array of holders.

    \item The success of photo-lithography depends on the choice of photo-resist that gets applied to the wafer. Photo-resist is a substance, that based on its state, may or may not pass photons of a specific wavelength through it. The first step of the process is to transfer the intended pattern to photo-resist and use it as a mask to etch the wafer. We have used a positive photo-resist solution AZ-12XT-10PR at a spin rate of 1500 rpm (this is the lowest possible spin rate to get even thickness) that would help us to achieve a thickness of ~12 um. The requirement of photo-resist thickness is defined by the etching time. For an etch time of $\sim$ 2.5 hours, the photo-resist is expected to get 8-10 micron deep holes.

    \item A lump of photo-resist is applied on the wafer and the wafer was rotated for 30 seconds. However, at this point, the photo-resist is liquid and hence mechanically not stable to hold its position. Once a uniform layer has formed, the photo-resist solution starts to harden due to the evaporation of the solvent thinner. This process is expedited by means of pre-exposure baking at 110\degree C for 180 seconds.  
    
    \item For the next step, the desired pattern is created on a transparent glass slide. The pattern consists of circular transparent spots on chromium coated opaque glass, which mimic the position and diameter of holes. Ideally, the spot centroid pattern of an MLA should be used to determine the location of the transparent patches for the corresponding mask. For general considerations here we have used regularly spaced circles. The chromium-plated area is opaque to the UV light. The remaining part of the wafer does not hinder the transmission of the UV light. 
    
    \item The wafer with the photo-resist layer is then carefully placed under the mask and held against it. The technique is called contact lithography. Then the mask-wafer assembly is illuminated by a 194 W UV light source for 30 seconds. The UV light transmits through to the transparent areas of the mask and chemically softens the photo-resist layer. However, the photo-resist layer beneath the opaque areas remains chemically hard. A post-exposure bake at 90\degree C for 60 seconds is used to mechanically harden the photo-resist.
    
    \item A developer solution AZ-300-MIF is used to dissolve the chemically soft parts of the photo-resist layer. The post-baked UV treated wafer is dipped in the developer solution and shaken for a minute (total) in two 30-second intervals. The UV-softened areas of the photo-resist layer get dissolved and holes are created on the photo-resist layer. The wafer is then washed with distilled water and dried, leading to the cleaning of the developer solution. 
\end{enumerate}

\subsection{Etching the wafer: the Bosch process}
\label{subsec:bosch}

\begin{figure}[H]
\centering
\includegraphics[width = \linewidth]{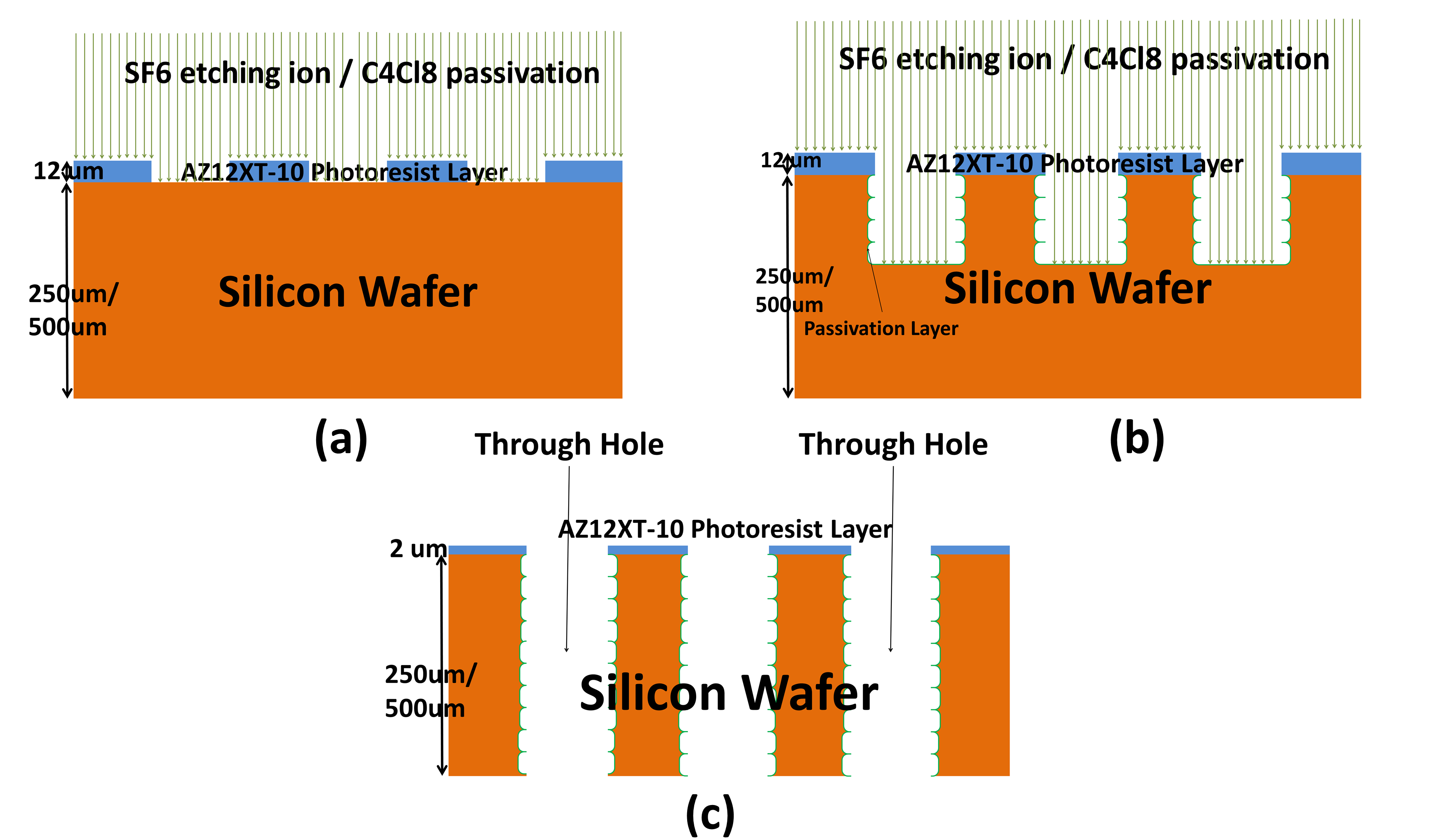}
\caption{Silicon wafer time-multiplexed deep etching process schematic. $SF_6$ is used as the etching reagent while $C_4Cl_8$ used as the passivating element.}
\label{fig:bosch}
\end{figure}

Deep reactive ion etching (DRIE), or the Bosch process (\citenum{drie}), of silicon, enables versatile and uniform micro-fabrication of high-aspect-ratio structures using the high etching rate of fluorine rich plasmas and the deposition of inhibiting films to obtain anisotropic profiles. An etching cycle, flowing only $SF_6$, is alternated with a sidewall-passivating cycle using only $C_4F_8$ (Figure \ref{fig:bosch}). The  $C_4F_8$ deposits a Teflon-like film (\citenum{driesilicon}) on the sidewalls to inhibit their etching during the subsequent $SF_6$ cycle. This passivating film is preferentially removed from the bottom of the trenches due to ion bombardment. Because of the alternating between etching and passivating cycles, DRIE is also -- perhaps more aptly -- referred to as time-multiplexed deep etching. The alternating etching and passivating cycles lead to scallops on the sidewalls of etched structures. The peak-to-valley height of these scallops can be controlled by operating conditions. 

Experimental studies (\citenum{driesilicon,tmde}) suggest that the achievable fillet radii, surface quality, and etch rate are functions of the etch conditions, namely the flow rate of $C_4F_8$ and $SF_6$, electrode power, chamber pressure, and etching cycle duration. Developing through-holes with vertical walls requires a careful combination of these parameters. The parameters we adopted are given in Table \ref{tab:bosch} are given by WCAM through a pre-installed recipe. However, we tweaked passivation and etching time, electrode power in the given prescription to suit our requirements. 

\begin{table}[h]
\centering
\caption{Bosch Deep Reactive Ion Etching Process Parameter Values for Fiber Holders}
\begin{tabular}{rrrl}

\textbf{Quantity}             & \textbf{Unit}   & \textbf{Value} & \textbf{Comment}                                                                                                         \\ \hline \hline
Pressure             & mTorr  & 20    & A variation of 17-21 is seen                                                                                    \\ \hline
Coil power           & Watt   & 640   &                                                                                                                 \\ \hline
Electrode power      & Watt   & 12    & Slightly higher value is also acceptable                                                                        \\ \hline
Bias                 & Volt   & 80    &                                                                                                                 \\ \hline
Passivation time     & second & 6     & Can be changed to 7s for thicker wafer                                                                          \\ \hline
Etching time         & second & 10    &                                                                                                                 \\ \hline
Passivator flow rate & sccm   & 96    &                                                                                                                 \\ \hline
Etchant flow rate    & sccm   & 101   & This is also accompanied by oxygen                                                                                 \\ \hline
Total process time   & minute & 150   & \begin{tabular}[c]{@{}c@{}}Total time may vary depending on\\ the thickness of photoresist coating\end{tabular} \\ \hline
\end{tabular}
\label{tab:bosch}
\end{table}

\subsection{Performance of Photolithography}

\begin{figure}[H]
\centering
\includegraphics[width = \linewidth]{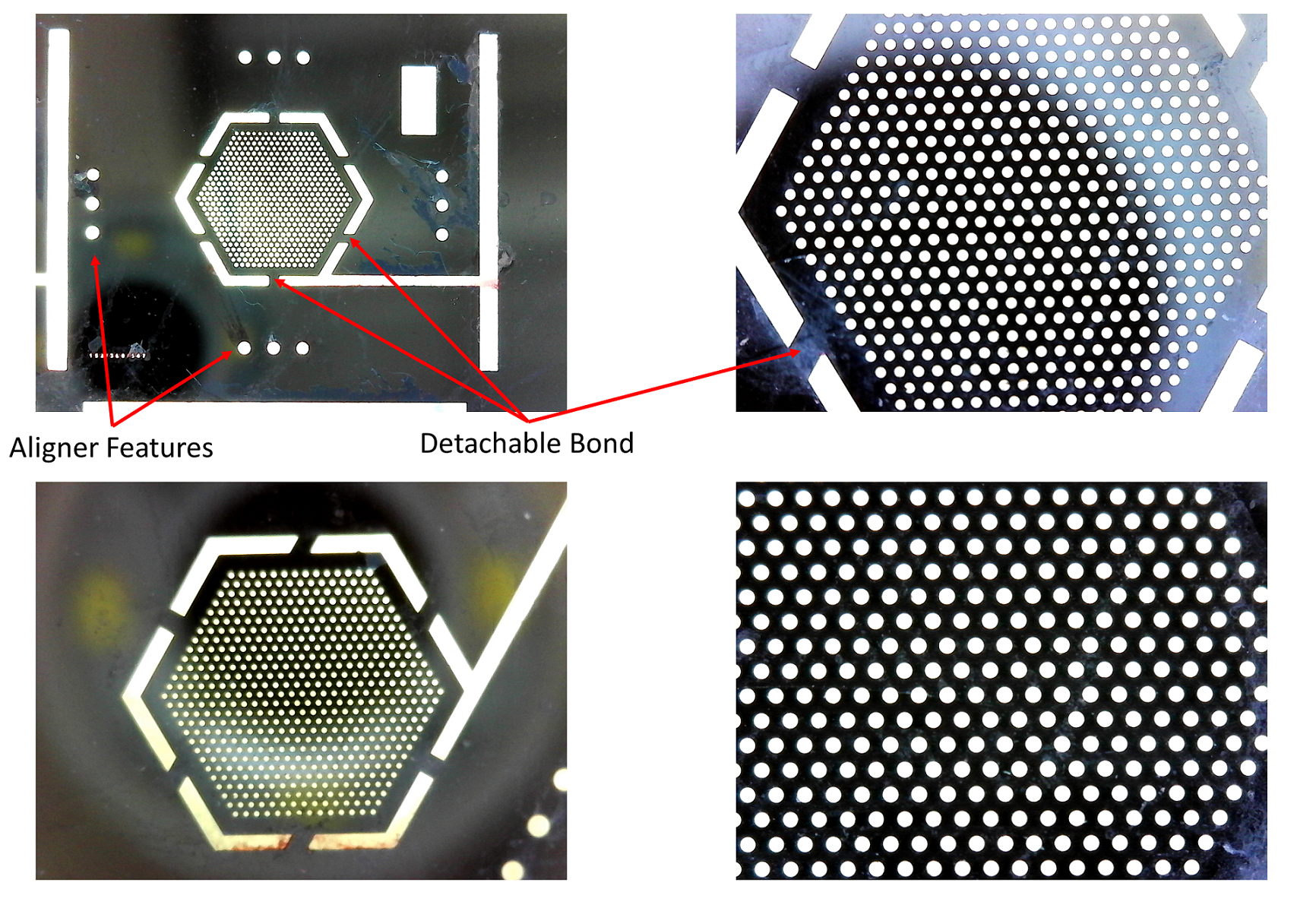}
\caption{Back illuminated image of a typical fiber array (250~$\mu$m thickness) manufactured at WCAM viewed at different magnification.}
\label{fig:lithores}
\end{figure}

Hole positions and their diameters of the fiber holders developed through photolithography have been measured from microscope images using a python script. The script takes a back-illuminated image of a hole array as input and fits an ellipse to each bright spot corresponding to a hole, as shown in Figure \ref{fig:lithores}. The fitted minor axis is defined as the hole diameter while the centroid of the ellipse is defined as the center of the hole. The measurement precision is driven by the signal to noise ratio (SNR) as well as the pixel dimension and array size of the imaging camera. We have kept the SNR high in order to measure the position with arbitrary precision. The imaging area is quite large (an 8~mm diameter area) while the USB microscope could only resolve up to 6.5 $\mu$m/pixel. We rotated the array multiple times and repeated the measurement procedure. The measurement error of diameter and position across different measurement is found to be $\pm$0.8 $\mu$m RMS.

\begin{figure}[H]
\centering
  \includegraphics[width=0.8\linewidth]{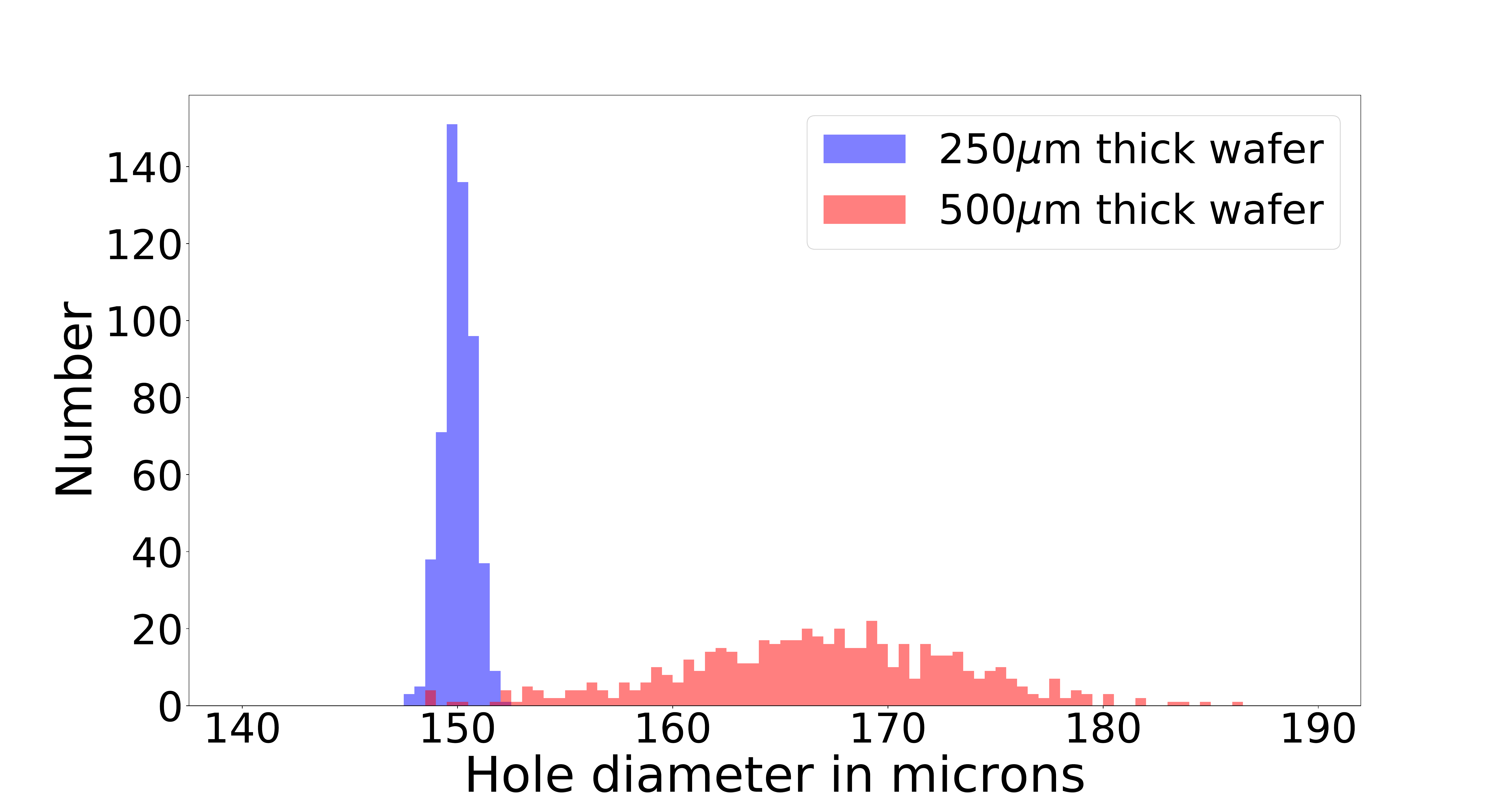}
  \caption{Error distribution in hole diameter for a typical mask from a 250 $\mu$m and a 500 $\mu$m wafer after performing photo-lithography at WCAM}
\label{fig:diameter}
\end{figure}

\begin{figure}[H]
\centering
  \includegraphics[width=0.8\linewidth]{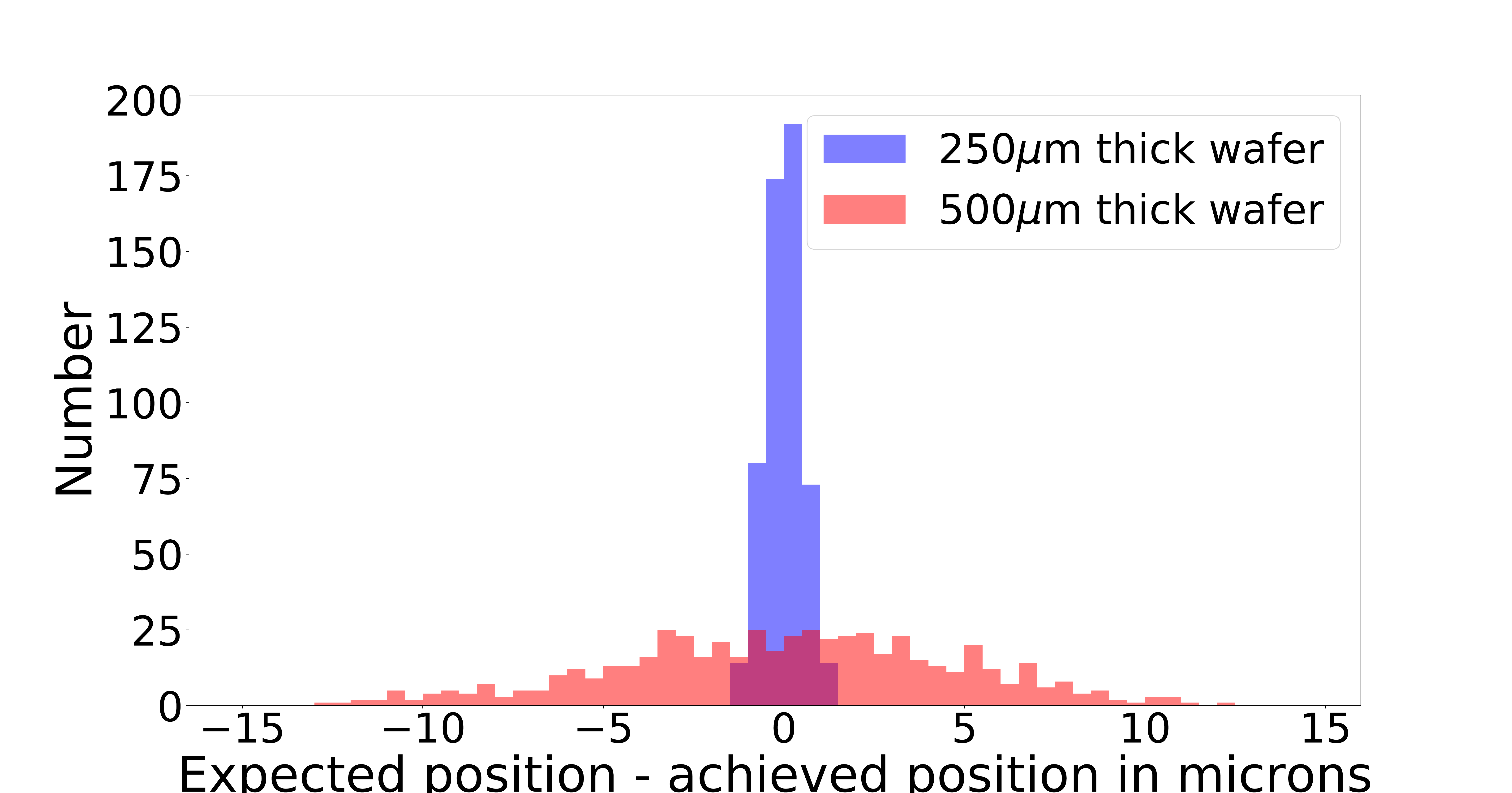}
\caption{Error distribution in hole relative position for a typical mask from a 250$\mu$m and a 500 $\mu$m wafer after performing photo-lithography at WCAM. }
\label{fig:position}
\end{figure}

We have etched two 500 $\mu$m and two 250 $\mu$m thick silicon wafers, each containing six hexagonal masks of 547 holes. The thicker wafers failed to yield consistent hole diameters and position as can be seen in Figure \ref{fig:diameter} \& \ref{fig:position}. This can be primarily attributed to the variation between silicon and photo-resist etch rate. For the given etch and passivation time (10 s and 6 s respectively) to achieve vertical walls, the used pressure level (90 mT) produced silicon to photo-resist etch ratio of $\sim$21. The etch ratio is defined as the ratio of silicon and photo-resist thickness etched over the same time scale. This implies we would require a photo-resist thickness of $\sim$ 24 $\mu$m to etch the 500$\mu$m wafer. Such photo-resist thickness is not achievable with the available photo-resist material. Other photo-resist material that can provide the required thickness can be used with metal only and not silicon. However, metal substrates cannot take advantage of the Bosch process (high accuracy hole diameter and position). While we could change the pressure to achieve a higher etch ratio, it would be very difficult to achieve etching-time and hole-diameter uniformity. Increasing the etch ratio would mean we wouldn't have sufficient control on the horizontal etching of walls which may lead to inaccurate hole diameters. We could also modulate the electrode power and coil power to reduce the photoresist etch rate and, in turn, increase the etch ratio. The required coil and electrode power ($<$400 W and $<$5 W respectively) would decrease the silicon etch rate thus reducing the etch ratio. This was found during the etching of 500 $\mu$m wafer: The photo-resist was etched out within 2.5 hrs while the silicon wafer etching did not get through the entire wafer thickness. Once the photo-resist was gone, the plasma would start etching the bare silicon. As a result, the hole diameter accuracy becomes unacceptable. 

\begin{figure}[H]
\centering

  \includegraphics[width=0.8\linewidth]{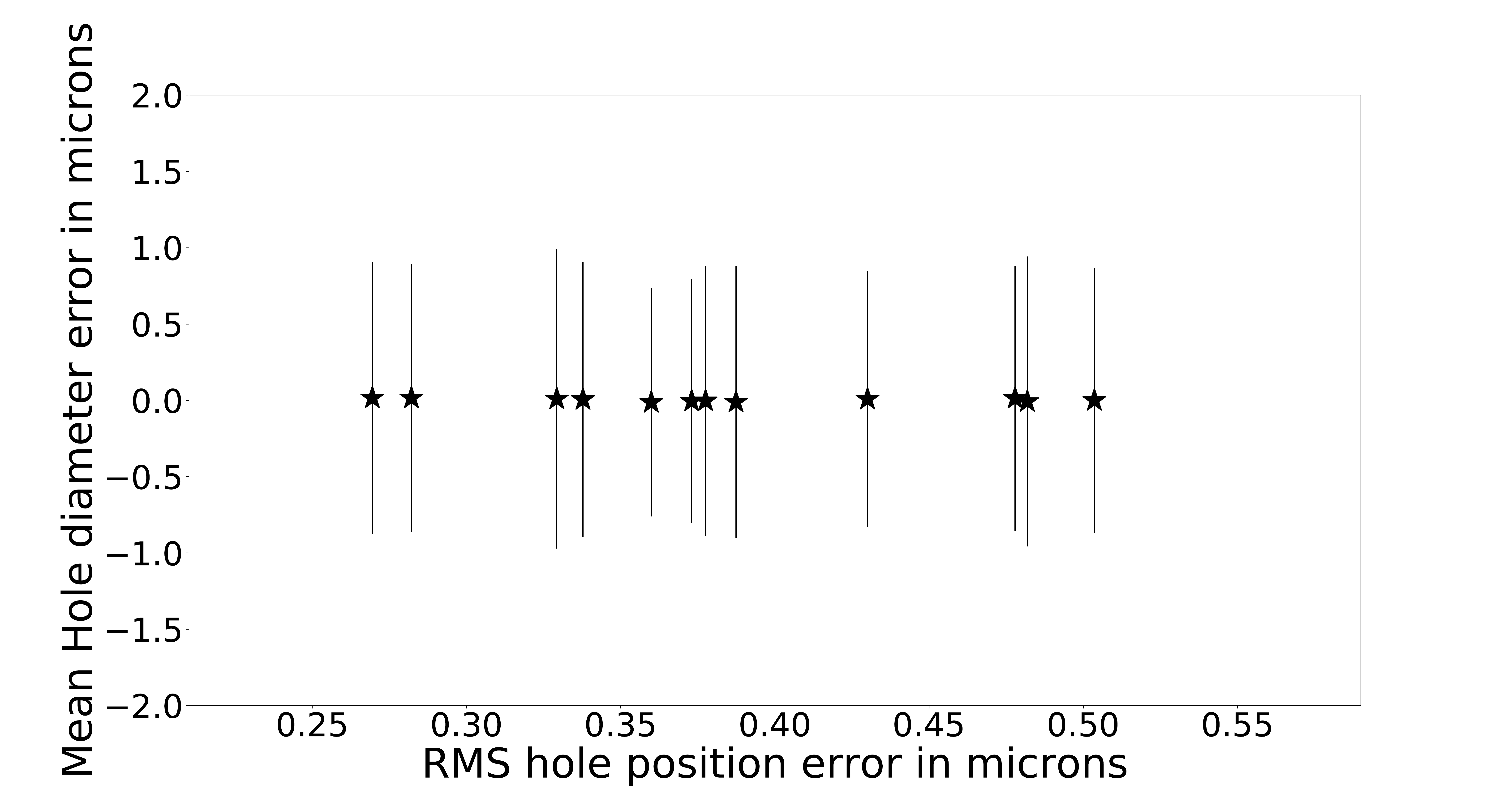}

\caption{Mean and peak to peak error in hole diameter against RMS relative position error for each of 12 masks from the two 250 $\mu$m wafers processed via photo-lithography at WCAM. }
\label{fig:errordist}
\end{figure}

\begin{figure}[H]
\centering
\includegraphics[width = 0.8\linewidth]{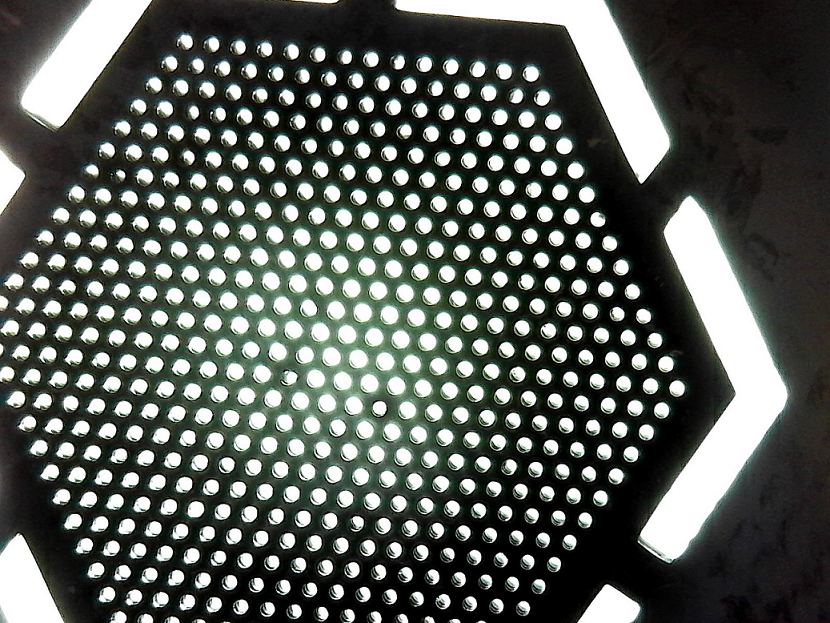}
\caption{Back-illuminated image of 12 stacked masks, each with 250~$\mu$m thickness, to produce a fiber-holder with a total thickness of 3~mm.}
\label{fig: full stack}
\end{figure}

Etching of 250 $\mu$m produced holders with 1 \& 0.5 $\mu$m RMS accuracy in hole diameter and relative position respectively, as shown in Figure \ref{fig:diameter} \& \ref{fig:position}. The recipe is repeatable and has a yield of 100\%. Since we need to stack 12 wafers to attain a thickness of 3 mm, additional features have been etched to align and hold the wafers to make the fiber insertion easier, as shown in Figure \ref{fig:litho}. We find that the mean and RMS error of diameters and positions do not vary across different masks on a wafer, as shown in Figure \ref{fig:errordist}. To demonstrate the suitability of these masks to achieve our thickness requirement, we have stacked all the wafers and found that they align together to form a clean hole for each fiber position, as shown in figure \ref{fig: full stack}.

\section{Results}
\label{sec:results}

\subsection{Optimized core filling}
%core filling optimization
\label{subsec:corefill}

\begin{figure}[H]
\centering
\includegraphics[width = 0.8\linewidth]{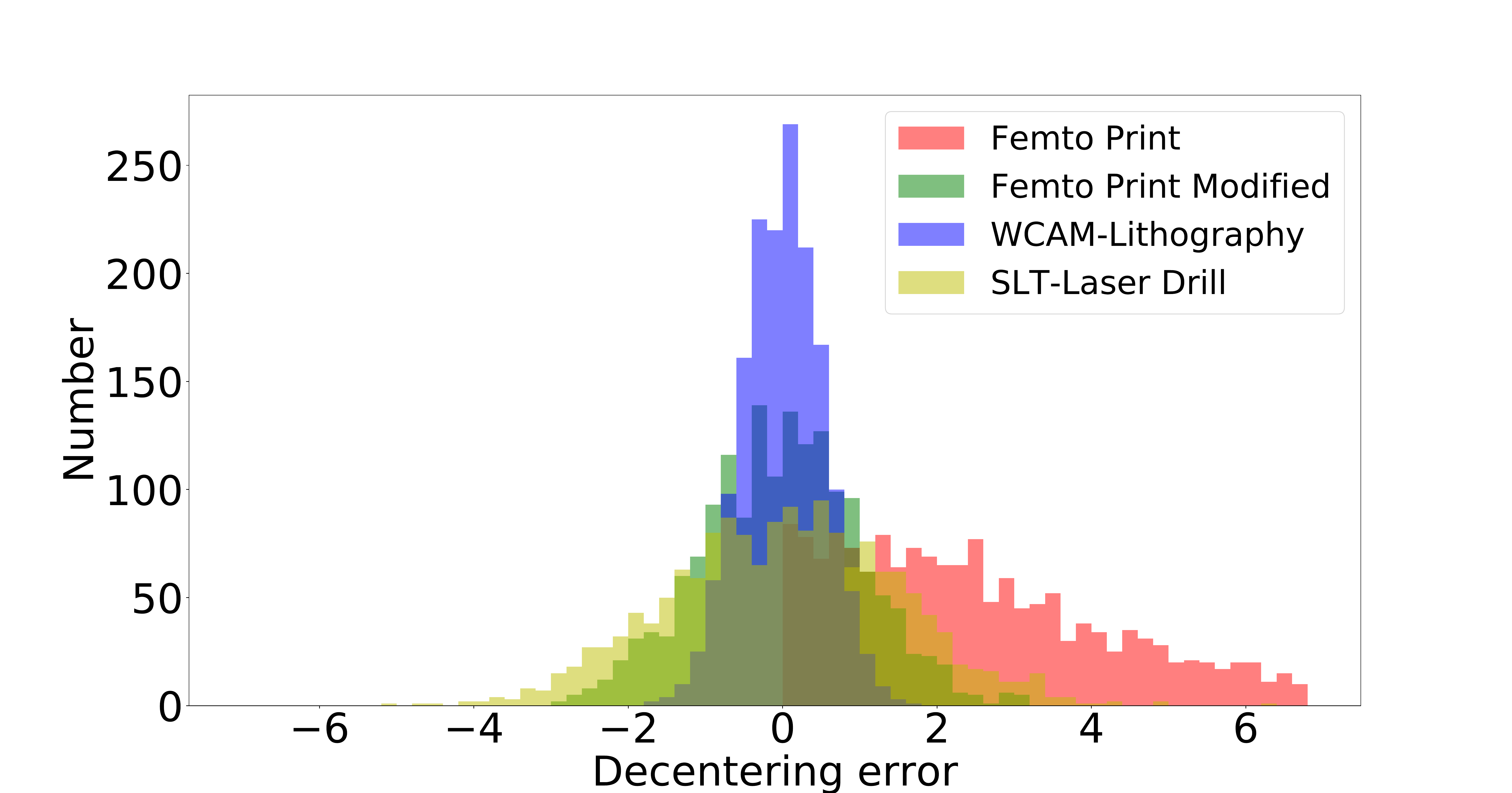}
\caption{Decenter distributions of the three fiber holder fabrication methods considered in this study (Sun-Light Tech, WCAM and FemtoPrint).}
\label{fig:distribution}
\end{figure}

\begin{figure}[H]
\centering
\includegraphics[width = 0.8\linewidth]{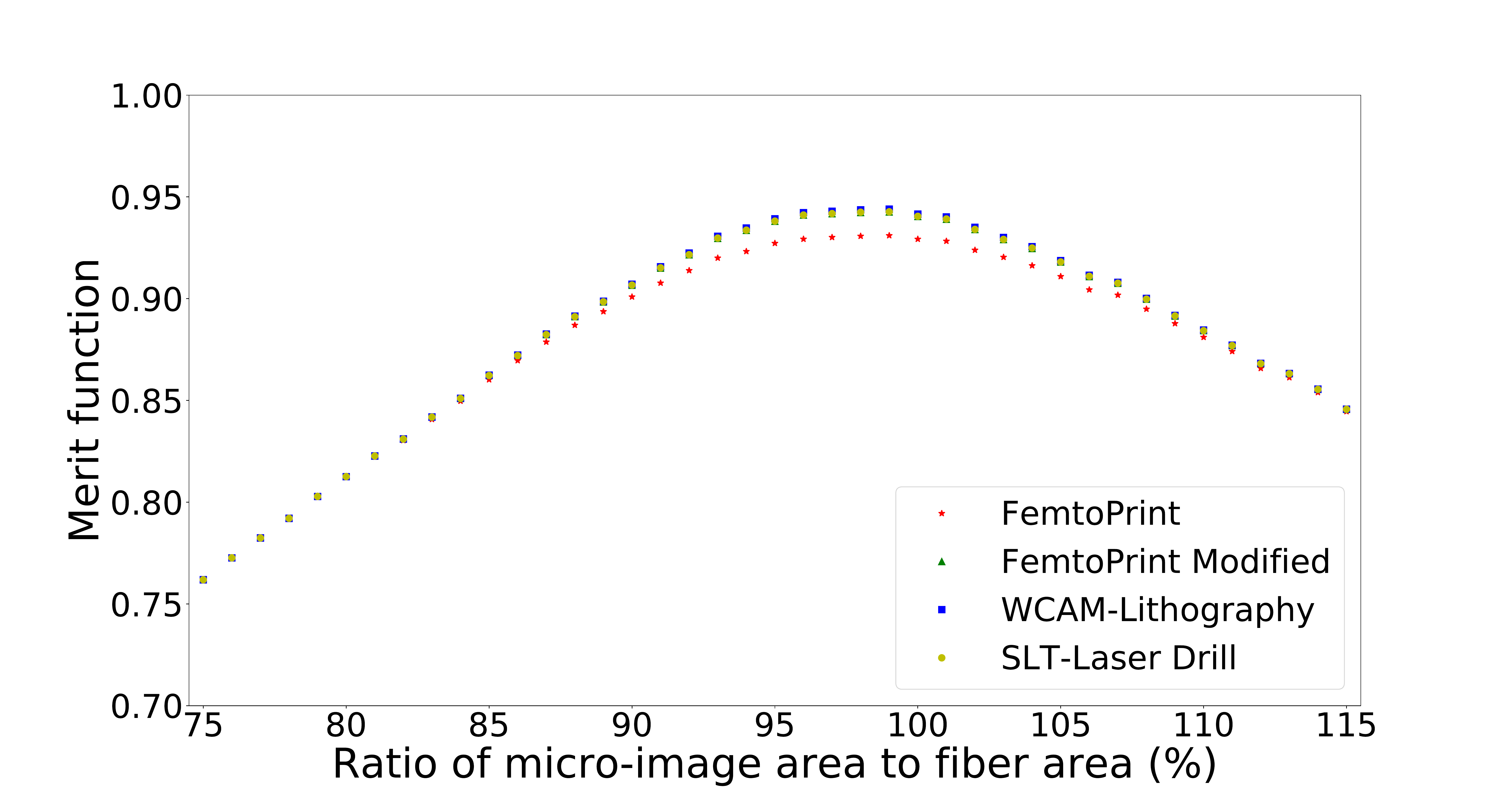}
\caption{Optimization of the fiber-core filling fraction (given as \%) based on our merit function for different hole positioning and diameter accuracies of three different fiber-holder fabrication methods (SLT, WCAM, and FP). The merit function assumes an ideal fiber with no focal ratio degradation, but this assumption only impacts the merit function amplitude not the location of the maximum with respect to the filling fraction. A core filling fraction of 97-98\% provides the least loss in terms of input photons and observing efficiency. This result is independent of the fiber core size as well as the input and output focal-ratios.}
\label{fig:meritfuction}
\end{figure}

We have generated three decenter distributions for a 2000 fiber array assuming Gaussian errors in positions with RMS values as described in subsection~\ref{subsec:schemes} for the three methods which are shown in Figure \ref{fig:distribution}. The number of spaxels is large enough to avoid small-sample bias in a gaussian distribution of random numbers. For each distribution, the IFU merit function $M_i$ is computed from Equation \ref{eq:finalmf} for a range of fiber-core input filling ($0.75 < A_i/A_c < 1.15$, where $A_c$ is the fiber core area). Given our instrument configuration defined in subsection~\ref{subsec:inst_config}, we adopt a fixed $\Omega_i$ corresponding to an input beam speed of f/4.2, and a fixed $A_o\Omega_o$ for a fully-illuminated fiber exit core with $A_o = A_c$ and f/4.2 output f-ratio (i.e. ideal fiber without focal ratio degradation). As shown in Figure \ref{fig:meritfuction}, the merit function increases with $A_i$ until $A_i/A_c$ approaches 97\%, beyond which the EE term in the merit function begins to dominate as light is lost outside of the over-filled fiber core. 

Different technologies differ in providing positioning precision. For example, FemtoPrint initially did not provide as good a solution as SLT or WCAM, but their improved precision is on par with the other techniques. A filling of $\sim$97-98\% of the fiber core is the optimal choice, delivering a peak in the merit function that is only 6\% lower than the ideal value (of unity). This optimum core filling fraction is independent of the number of spaxels (as the merit function is an average); fiber core size (as the fill fraction is a ratio of areas); and input or output focal-ratio. The {\it amplitude} of the merit function {\it does} depend on the ratio of the input-to-output f-ratios; here we have assumed the case without FRD such that the input and output f-ratios are equal. We have no evidence that FRD depends on the core filling fraction. We assume a contiguous distribution of microlenses for the best possible application of the microlens-coupled fiber IFU technique.

\begin{figure}[H]
\centering
\includegraphics[width = 0.8\linewidth]{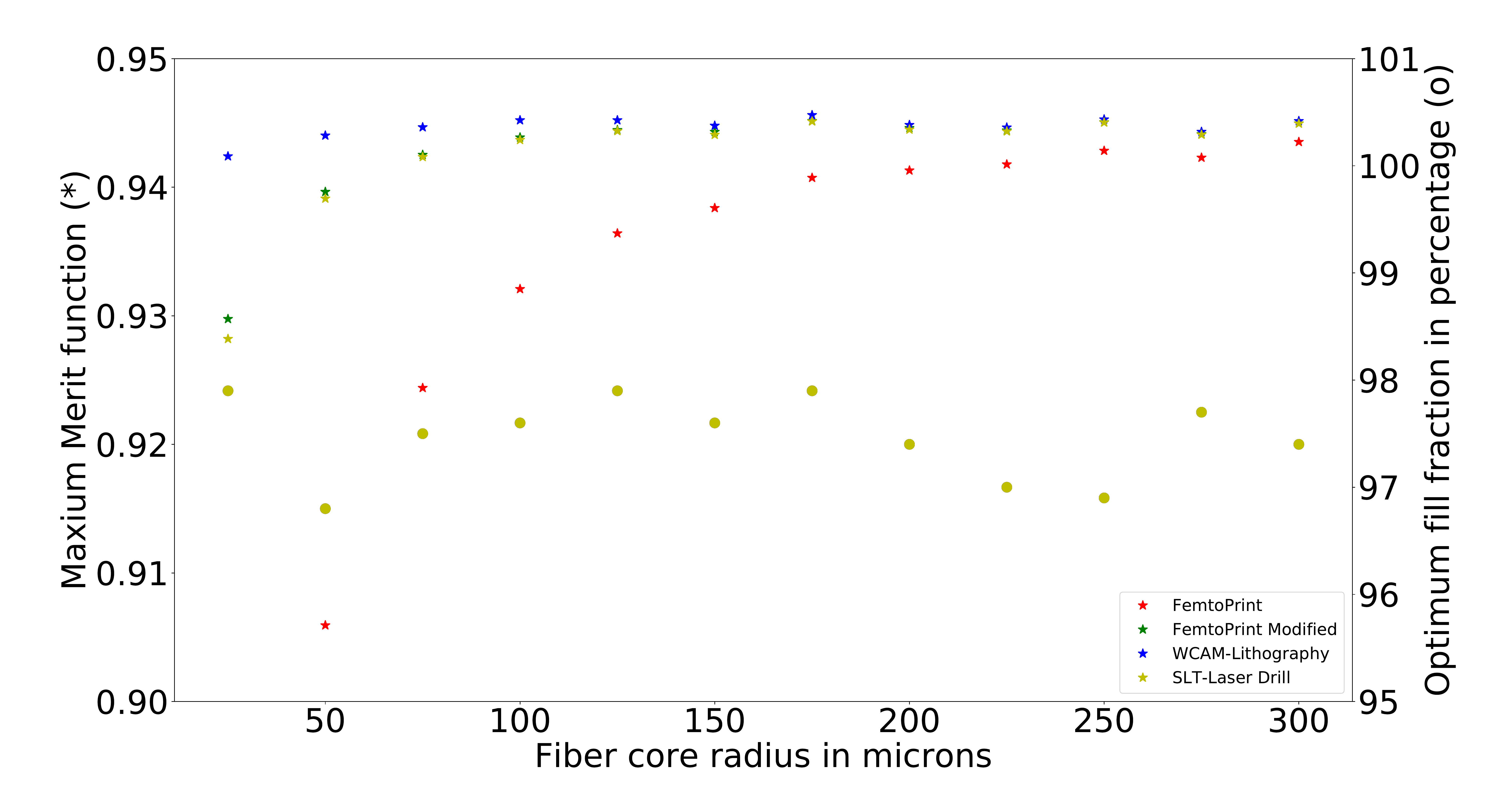}
\caption{Variation of maximum achievable merit function and optimum fill fraction against various fiber core dimension for different techniques of fiber holder development in an ideal fiber having no focal ratio degradation. The optimum core fill fraction is same for all the techniques given a fixed core size and varies around 97-98\% across all core sizes.}
\label{fig:MF vs core}
\end{figure}

It is also interesting to note that although SLT/Modified FP hole positioning accuracy is lower than WCAM-lithography, we are not losing much in terms of the merit function maximum value. This primarily stems from the fact that the difference between input spot and fiber core diameters remains comparable to the fiber decenter distribution for SLT and modified FP methods; by increasing the fiber positioning we would gain neither in throughput nor in observing efficiency for a core diameter of 100 $\mu$m. However, changing the core size would change this scenario as shown in figure \ref{fig:MF vs core} which describes the variation of maximum achievable merit function and optimum fill fraction from different techniques against core diameter. The optimum fill fraction for a given fiber core radius remains the same across all techniques and also distributed around 97-98\% for different core sizes. For the smaller fibers ($\leq$ 100$\mu$m radius), the difference between core and micro image diameter would be equivalent to decenter for WCAM but much smaller for modified FP and SLT. As the core diameter grows, for the same fill fraction, the difference gets higher and eventually becomes equivalent to the decenter distribution of SLT \& modified FP and at this point, they perform very similar to WCAM.

\subsection{Fiber holder thickness requirements}

\begin{figure}[H]
\centering
\includegraphics[width = 0.8\linewidth]{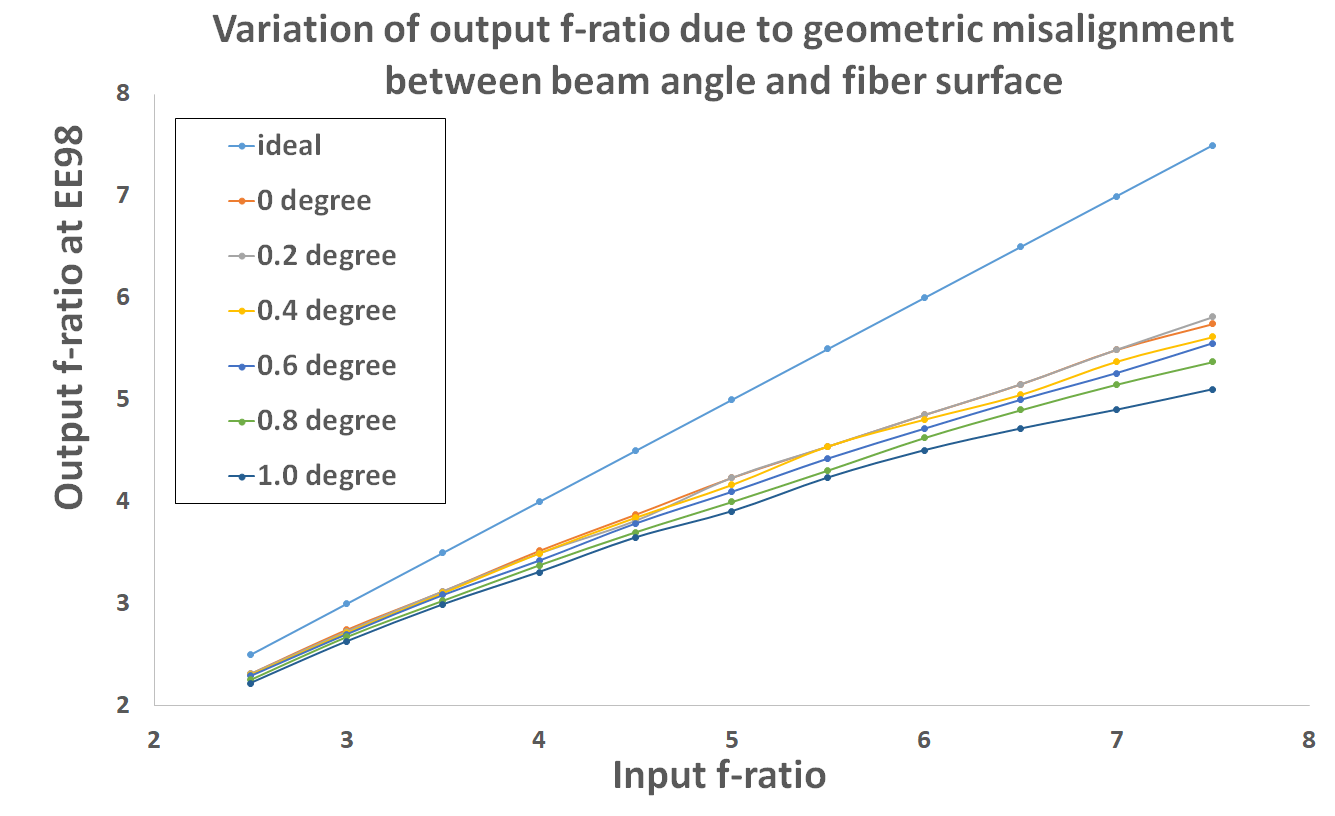}
\caption{Effect of misalignment between fiber face and optical beam (termed as tilt) on the output focal ratio without incorporating the optically introduced focal ratio degradation.}
\label{fig:inout}
\end{figure}

\begin{figure}[H]
\centering
\includegraphics[width = 0.8\linewidth]{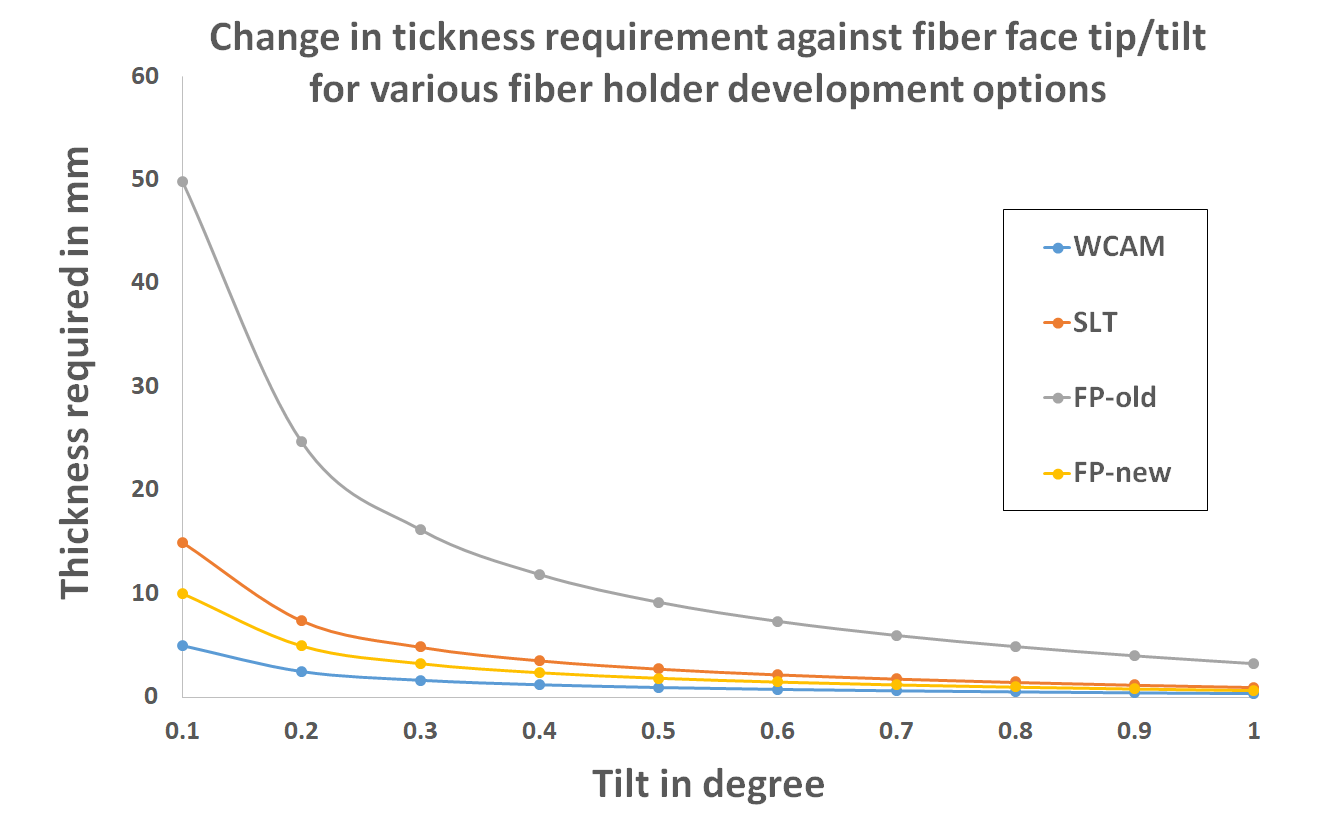}
\caption{Variation in fiber holder thickness requirement for different tilt angles and different holder development techniques. }
\label{fig:thicknessvsTilt}
\end{figure}

In subsection \ref{subsec:thickness} we have described the thickness required for the fiber holder that would hold the fiber array for the SMI instrument for the SALT. Given the three available techniques, here we present a generic treatment towards defining the required thickness of a fiber holder. Figure \ref{fig:inout} shows the variation of output focal ratio due to misalignment between the input beam and the fiber face. This tip/tilt can be converted to a thickness requirement based on the fiber hole diameter accuracy of different fiber holder development techniques as provided in subsection \ref{subsec:schemes}. The change in the thickness required to achieve a range of tilt angle has been shown in figure \ref{fig:thicknessvsTilt}. At a given tilt tolerance the thickness requirement decreases with increasing hole diameter accuracy. For a tilt of 0.2$\degree$ WCAM needs the smallest thickness of 2.5 mm. At a faster input focal ratio ($<$f/3.5), the optomechanical design of the IFU can accept tilts up to 1 degree at which all of the holder development techniques provide equivalent thickness requirement ($<$1 mm).

\section{Summary}
\label{sec:summary}

Developing multi-spaxel microlens-fiber coupled IFUs requires careful alignment within $\sim$ 1\% of the fiber diameter. Over or underfilling the fiber core with the micro-image can lead to severe light-loss. We find 97-98\% is the ideal filling fraction of the fiber core with the micro-image/micro-pupil for a fiber-positioning accuracy of \textless 1 $\mu$m RMS. In addition to positioning accuracy, the fiber holder must minimize the tilt between fiber and the microlens to minimize geometric FRD. An acceptable tilt of $0.1\degree$ will ensure f/4 output for an input beam of f/4.2 at 98\% encircled energy. Several methods of fiber holder development have been pursued and the effect of their drilling uncertainties on throughput has been analyzed. It is found that FemtoPrint, Sun-Light Tech. and WCAM can be used with equal efficiency. The process of photo-lithography (WCAM) has been deployed to etch 250~$\mu$m and 500~$\mu$m thick silicon wafers. The recipe was successful for the thinner wafer, fulfilling the desired hole position and diameter accuracy of $\sim$0.5 and 1 $\mu$m RMS respectively with a yield of 100\%.

The three possible techniques can be used to achieve similar results but at different costs. FemtoPrint (FP) can produce holders of 5 mm thickness and thus stacking errors are not present but we found that this is the costliest among the three techniques. Fiber holders manufactured at Sun-Light Tech (SLT) may or may not require stacking depending on the fiber size as the hole diameter to drill thickness is fixed (1:10); individual SLT plates are lower cost than FemtoPrint. Holders fabricated at WCAM must be stacked as the individual wafer thickness can only be 250 $\mu$m. The stacking error may be mitigated with aligning features and the process cost is significantly cheaper than the other two. However, for WCAM, labor costs brings the total cost much closer to the other techniques for a small production set. An instrument requiring several IFUs (similar to SMIFU specification) may find WCAM costs to be competitive with the other techniques since the labor does not increase significantly as the process of photolithography can be made parallel (up to 3 wafers with the current facility) as well as multiple wafers can be etched simultaneously. It is difficult to generalize the relative costs of the different method since they depend on the specific application, but we conclude below with the relative costs for our application and indicate how these relative costs change depending fiber and array size as well as tilt requirement.

For our nominal case requiring $<0.1\degree$ fiber tilt for 200~$\mu$m core fibers, leading to 3~mm holder thickness, WCAM lithography is about 10\% cheaper than SLT and a factor of 4 less than FP. Keeping the number of fibers and the tilt requirement fixed, for larger fiber diameter ($>$500~$\mu$m) SLT becomes the most cost-effective approach, followed by WCAM and then FP. This change in relative cost comes about because SLT's technique is limited by aspect ratio, and this becomes salient for large fiber sizes. For stricter tilt requirement this trend continues. On the other hand for large fiber diameters ($>$500~$\mu$m) and loose tilt criteria ($>0.1\degree$) FP becomes cheaper followed by WCAM and SLT. At small fiber diameters ($<$100~$\mu$m) and tight tilt criteria ($<0.025\degree$) FP and SLT are comparable but remain factors of a few more expensive than WCAM. 

Overall, we were impressed with FemtoPrint's capabilities and performance, SLT's performance and low cost, and WCAM's facilities for in-house custom-development. For instrument projects without in-house access to superb photolithorgraphic facilities such as WCAM, or without personnel trained to use such facilities, these commercial vendors, or ones like them, should serve as essential for successful, higher-performance fiber-lenslet coupling.

\acknowledgments 
This research was supported by funds from the University of Wisconsin-Madison Graduate School, NSF AST-1517006. The authors gratefully acknowledge the use of facilities and instrumentation at the UW-Madison Wisconsin Centers for Nanoscale Technology (wcnt.wisc.edu) partially supported by the NSF through the University of Wisconsin Materials Research Science and Engineering Center (DMR-1720415). We also thank Frank Pfefferkorn for his input towards vendor choice and fabrication options.

\bibliography{references}   % bibliography data in report.bib
\bibliographystyle{spiejour}   % makes bibtex use spiejour.bst

\end{document}